\documentclass{article}

\usepackage{arxiv}

\usepackage[utf8]{inputenc} 
\usepackage[T1]{fontenc}    
\usepackage{hyperref}       
\usepackage{url}            
\usepackage{booktabs}       
\usepackage{amsfonts}       
\usepackage{nicefrac}       
\usepackage{microtype}      
\usepackage{lipsum}		
\usepackage{graphicx}
\usepackage{doi}
\usepackage{adjustbox}

\usepackage{subcaption}
\usepackage{gensymb}
\usepackage{mathtools}
\usepackage{siunitx} 
\usepackage{multirow}

\title{NeCA: 3D Coronary Artery Tree Reconstruction from Two 2D Projections via Neural Implicit Representation}

\author{Yiying Wang $^{1}$, Abhirup Banerjee $^{1,2}$* and Vicente Grau $^{1}$
\\
$^{1}$ \quad Institute of Biomedical Engineering, Department of Engineering Science, University of Oxford\\
$^{2}$ \quad Division of Cardiovascular Medicine, Radcliffe Department of Medicine, University of Oxford\\
* Correspondence: abhirup.banerjee@eng.ox.ac.uk	
}
\date{}

\begin{document}
\maketitle

\begin{abstract}
Cardiovascular diseases (CVDs) are the most common health threats worldwide. 2D X-ray invasive coronary angiography (ICA) remains the most widely adopted imaging modality for CVD assessment during real-time cardiac interventions. However, it is often difficult for cardiologists to interpret the 3D geometry of coronary vessels based on 2D planes. Moreover, due to the radiation limit, often only two angiographic projections are acquired, providing limited information of the vessel geometry and necessitating 3D coronary tree reconstruction based only on two ICA projections. In this paper, we propose a self-supervised deep learning method called NeCA, which is based on neural implicit representation using the multiresolution hash encoder and differentiable cone-beam forward projector layer, in order to achieve 3D coronary artery tree reconstruction from two 2D projections. We validate our method using six different metrics on a dataset generated from coronary computed tomography angiography of right coronary artery and left anterior descending artery. The evaluation results demonstrate that our NeCA method, without requiring 3D ground truth for supervision or large datasets for training, achieves promising performance in both vessel topology and branch-connectivity preservation compared to the supervised deep learning model. The code to our work is available at \href{https://github.com/WangStephen/NeCA}{our GitHub repository}.
\end{abstract}


\section{Introduction}
Cardiovascular diseases (CVDs) are the most common cause of death worldwide \cite{who_2021}. X-ray invasive coronary angiography (ICA) remains the most widely adopted imaging modality for CVD assessment during real-time cardiac interventions \cite{lashgaripatient}. ICA acquires 2D projections of the coronary tree, which makes it difficult for cardiologists in clinical practice to understand the global vascular anatomical structure due to vessel overlap and foreshortening. Moreover, potential adverse effects of the higher amount of radiographic contrast agent and higher radiation required for long-time exposure to X-rays restrict the number of angiographic projections acquired; typically, 2--5 projections are acquired, providing limited information of the vessel structures. Therefore, it is of great significance to perform 3D coronary tree reconstruction from only two 2D projections to provide spatial vascular information, which can significantly reduce the risks of subjective interpretation of the 3D coronary vasculature from 2D views and decrease the complexity of interventional surgeries. 

\par Several conventional mathematical methods have been proposed for 3D coronary tree reconstruction from ICA projections \cite{ccimen2016reconstruction,8864089,Banerjee2018stacom,Banerjee2019premi}, but they usually depend on traditional stereo-vision algorithms, requiring substantial manual interactions. The emergence and prosperity of deep neural networks have enabled 3D automated reconstruction from limited views in medical images \cite{wang2023review,ratul2021ccx,9176040}. Most of them need large training datasets and work in a supervised learning manner, but the acquisition of paired data has always been a challenge in real clinics. Recently, Neural Radiance Fields (NeRF) \cite{gao2022nerf} have made a significant contribution to the field of computer vision, allowing for neural implicit representation and novel view synthesis. In neural implicit representation learning, a bounded scene is parameterised by a neural network as a continuous function that maps spatial coordinates to metrics such as occupancy and colour. The optimization of NeRF only relies on several images from different viewpoints. Based on NeRF, Neural Attenuation Fields \cite{zha2022naf} (NAFs) are proposed to tackle the problem of sparse-view cone-beam computed tomography (CT) reconstruction, which require at least 50 projections. \cite{shen2022nerp} proposed a neural implicit representation learning methodology to reconstruct CT images, which performs on 10, 20, and 30 projections.

\par Few studies have explored deep learning for 3D vessel reconstruction from limited projections. Reconstructing 3D cerebral vessels using deep learning has received some attention in recent years. A self-supervised learning model \cite{zhao2022self} was proposed for the 3D reconstruction of cerebral vessels based on ultra-sparse X-ray projections. \cite{zuo20212d} implemented an adversarial network for 3D neurovascular reconstruction based on biplane angiograms, but the results are limited, with flaws occurring near crossed vessels. Some deep learning-based studies also attempted 3D coronary tree reconstruction from limited projections. \cite{wang2020weakly} used coronary computed tomography angiography (CCTA) data to simulate projections and trained a weakly supervised adversarial learning model for 3D reconstruction from two projections. However, their model requires large training datasets (8800 data in the experiments), with the 3D ground truth used in the discriminator. \cite{wang2024deep} also used a large CCTA dataset to simulate projections for training. \cite{ibrahim20223d,uluhan20223d,iyer2023multi} generated 3D synthetic coronary tree data and simulated corresponding 2D projections to train supervised learning models; their models require more than two projections for training. \cite{bransby20233d} used bi-planar ICA data to reconstruct a single coronary tree branch in a supervised learning setup. \cite{maas2023nerf} proposed a NeRF-based model to achieve 3D coronary tree reconstruction from limited projections without involving 3D ground truth in training. However, they tested the performance only on two 3D studies, and the number of required projections is at least four. Despite the improvement in deep neural networks, 3D coronary tree reconstruction from two projections without involving corresponding 3D ground truth and large training datasets remains challenging.

\par In this paper, we propose a self-supervised deep learning method named NeCA, which is based on neural implicit representation to achieve 3D coronary artery tree reconstruction from only two projections. Our method requires neither 3D ground truth for supervision nor large training datasets. It iteratively optimises the reconstruction results in a self-supervised fashion with only the projection data of one subject as input. Our proposed method utilizes the advantages of the multiresolution hash encoder \cite{muller2022instant} to encode point coordinates, residual multilayer perceptrons (MLP) to predict point occupancy, and a differentiable cone-beam forward projector layer \cite{jonas_adler_2017_249479} to simulate projections. The simulated projections are then learned from the input projections by minimising the projection error in a self-supervised manner. Our method aims to learn and optimise the neural representation for the entire image and can directly reconstruct the target image by incorporating the forward model of the imaging system. We use a public CCTA dataset \cite{ZENG2023102287} to validate our model's feasibility on the task based on six metrics. The evaluation results indicate that our proposed NeCA model, without 3D ground truth for supervision or large datasets for training, achieves promising performance in both vessel topology preservation and maintaining branch connectivity compared to an equivalent supervised learning model. The main contributions of this work are:
\begin{enumerate}
\item \textbf{3D coronary tree reconstruction using self-supervised learning from only two projections:} Our proposed deep learning method achieves 3D coronary artery tree reconstruction from two projections where neither 3D ground truth for supervision nor large training datasets are required.
\item \textbf{Neural implicit representation learning:} We leverage the advantages of MLP neural networks as a continuous function to represent the coronary tree in 3D space in order to enable mapping from encoded coordinates to corresponding occupancies.
\item \textbf{The applications of multiresolution hash encoder and differentiable cone-beam forward projector layer:} We combine a learnable hash encoder and a differentiable projector layer in our model to allow for efficient feature encoding and self-supervised learning from 2D input projections.
\item \textbf{Evaluations:} We perform thorough evaluation of our model on the right coronary artery and left anterior descending artery in terms of six quantitative metrics.
\end{enumerate}

\section{Materials and Methods}
\subsection{Dataset} 
\begin{figure}[!h]
     RCA
	\begin{subfigure}[H]{0.2\textwidth}
         \centering
         \includegraphics[width=\textwidth]{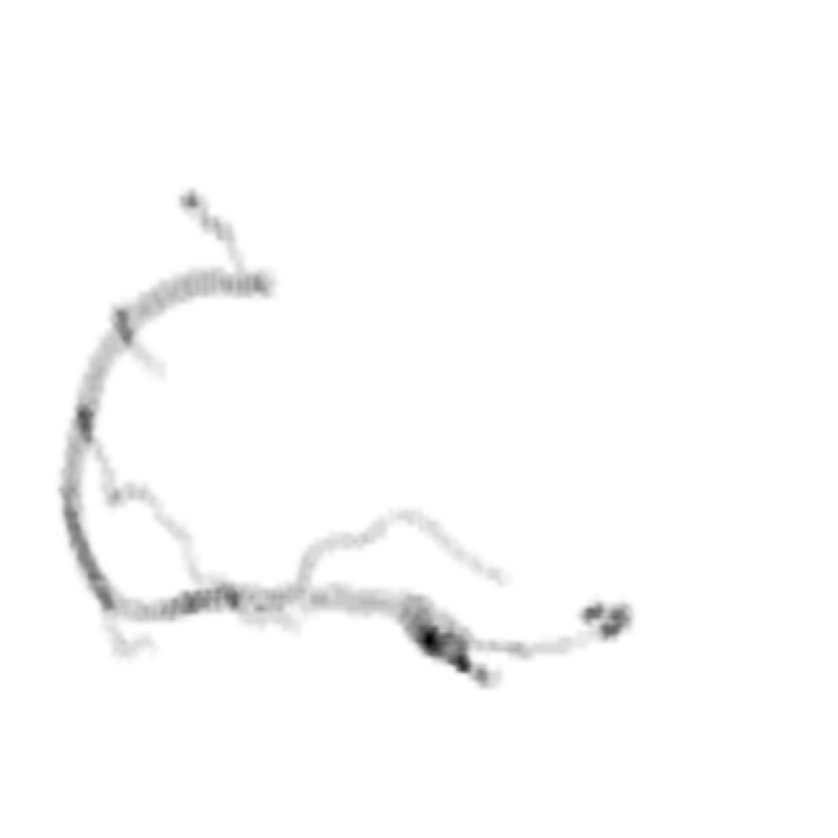}
         \caption*{1st projection}
     \end{subfigure}
     \hfill
	\begin{subfigure}[H]{0.2\textwidth}
         \centering
         \includegraphics[width=\textwidth]{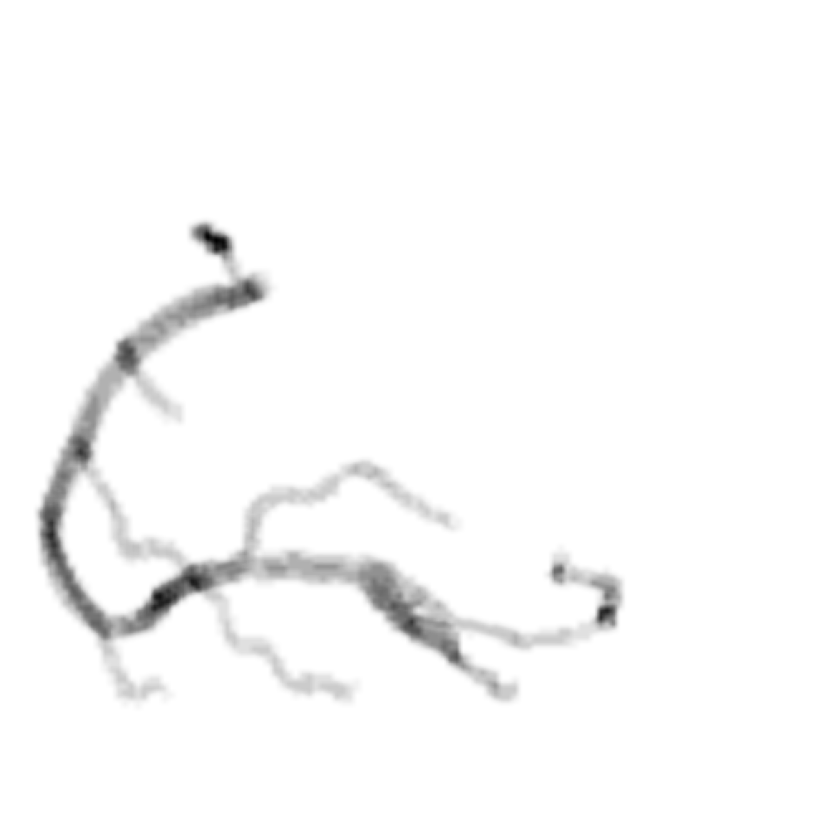}
         \caption*{2nd projection}
     \end{subfigure}
     \hfill
     LAD
	\begin{subfigure}[H]{0.2\textwidth}
         \centering
         \includegraphics[width=\textwidth]{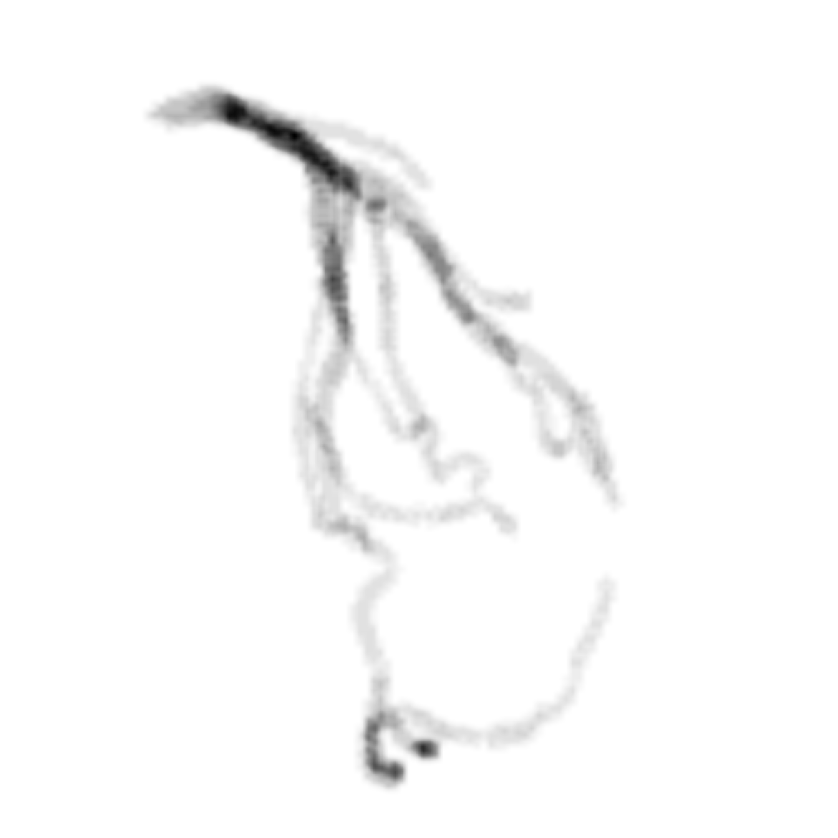}
         \caption*{1st projection}
     \end{subfigure}
     \hfill
	\begin{subfigure}[H]{0.2\textwidth}
         \centering
         \includegraphics[width=\textwidth]{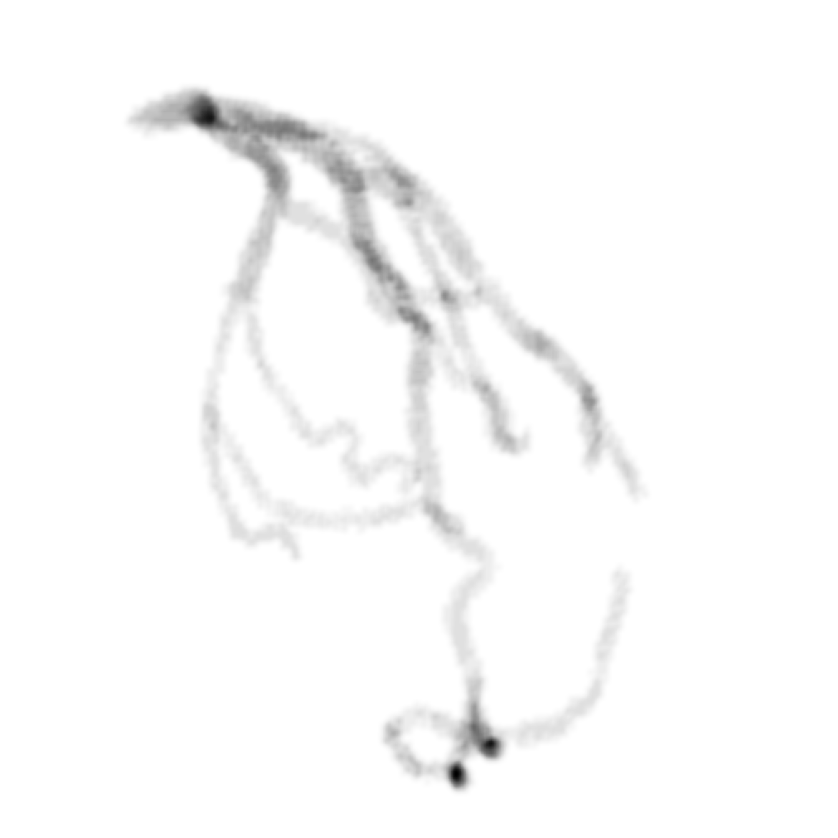}
         \caption*{2nd projection}
     \end{subfigure}
    \caption{An example of two projections generated from RCA and LAD data.}
        \label{projections_example}
\end{figure}

We use a public CCTA dataset \cite{ZENG2023102287} containing binary segmented coronary trees for our study, splitting the coronary trees into the right coronary artery (RCA) and left anterior descending (LAD) artery. Since our model is an optimization-based method for each individual data point, we do not need training/validation split. We use 67 RCA data and 79 LAD data points as the test set. We perform cone-beam forward projections on the CCTA data to generate the input projections with simulated attenuated X-ray intensities based on the Operator Discretization Library (ODL) \cite{jonas_adler_2017_249479}. For each CCTA data point, we generate only two projections to use in our model for 3D coronary tree reconstruction. The projection geometries for RCA and LAD are illustrated in Table~\ref{proj_geo}, which mimic the ones generally used in clinics. Figure~\ref{projections_example} illustrates an example of two projections generated from both RCA and LAD.
\begin{table}[!h]
\caption{The projection geometry to simulate cone-beam forward projections for both RCA and LAD. DSD: distance for source to detector; DSO: distance for source to origin.}\label{proj_geo}
\centering 
\begin{tabular}{clcc}
\toprule
\textbf{Data}                                                            & \multicolumn{1}{c}{\textbf{Geometry}} & \multicolumn{1}{c}{\textbf{First Projection Plane}} & \textbf{Second Projection Plane}     \\ \midrule
\multirow{5}{*}{RCA and LAD} & Detector spacing                       & \multicolumn{2}{c}{$0.2769 \times 0.2769$ $mm^2$ to $0.2789 \times 0.2789$ $mm^2$}         \\ \cmidrule{2-4} 
                                                                         & Detector size                          & \multicolumn{2}{c}{$512 \times 512$}                                                       \\ \cmidrule{2-4} 
                                                                         & Volume spacing                         & \multicolumn{2}{c}{$90 \times 90 \times 90$ $mm^3$ to $105 \times 105 \times 105$ $mm^3$}  \\ \cmidrule{2-4} 
                                                                         & Volume size                            & \multicolumn{2}{c}{$128 \times 128 \times 128$}                                            \\ \midrule
\multirow{5}{*}{RCA}                                                     & DSD                                    & \multicolumn{1}{c}{970 $mm$ to 1010 $mm$}           & 1050 $mm$ to 1070 $mm$               \\ \cmidrule{2-4} 
                                                                         & DSO                                    & \multicolumn{1}{c}{745 $mm$ to 785 $mm$}            & $\pm 3$ $mm$ to the 1st projection \\ \cmidrule{2-4} 
                                                                         & Primary angle                          & \multicolumn{1}{c}{$18 \degree$ to $42 \degree$}    & $-8 \degree$ to $8 \degree$          \\ \cmidrule{2-4} 
                                                                         & Secondary angle                        & \multicolumn{1}{c}{$-8 \degree$ to $8 \degree$}     & $18 \degree$ to $42 \degree$         \\ \midrule
\multirow{5}{*}{LAD}                                                     & DSD                                    & \multicolumn{1}{c}{1030 $mm$ to 1090 $mm$}          & $+ 70$ $mm$ to the 1st projection  \\ \cmidrule{2-4} 
                                                                         & DSO                                    & \multicolumn{1}{c}{740 $mm$ to 760 $mm$}            & $+ 3$ $mm$ to the 1st projection   \\ \cmidrule{2-4} 
                                                                         & Primary angle                          & \multicolumn{1}{c}{$-8 \degree$ to $8 \degree$}     & $-47 \degree$ to $-23 \degree$       \\ \cmidrule{2-4} 
                                                                         & Secondary angle                        & \multicolumn{1}{c}{$18 \degree$ to $42 \degree$}    & $21 \degree$ to $45 \degree$         \\ \bottomrule
\end{tabular}
\end{table}

\subsection{Proposed Model}
Our proposed model NeCA consists of five stages and allows for end-to-end learning. First, we normalise the coordinate index in the image spatial field according to resolution. Then, for each voxel point, we use a multiresolution hash encoder \cite{muller2022instant} to encode their normalized coordinates to obtain the corresponding multiresolution spatial feature vectors. These feature vectors are next sent to the residual MLP to predict the occupancy at the position of that point. The occupancy predictions of all the points form the 3D coronary tree reconstruction results. After that, we simulate the X-ray forward projections from the 3D predicted reconstruction based on the projection geometry of the input. Finally, these simulated projections are learned iteratively against the input projections in a self-supervised way. Stages 2 to 5 of our proposed model are illustrated in Figure~\ref{model}.

\subsubsection{Coordinate Normalization}
The input to the model is a set of integer coordinates ${\mathbf x = (x,y,z)}$ based on the number of voxels $n_{vx} \times n_{vy} \times n_{vz}$ in 3D volume ranging in $(1 \text{ to } n_{vx},1 \text{ to } n_{vy},1 \text{ to } n_{vz})$. We normalise the coordinates from these voxels according to the voxel spacing $s_{vx,vy,vz}$ along each axis, as calculated in Equation~(\ref{eq_norm}). These normalized coordinates ${\mathbf x' = (x',y',z')}$ are then sent to a multiresolution hash encoder at the next stage to efficiently obtain the corresponding spatial feature vectors.
\begin{equation}\label{eq_norm}
\begin{gathered}
n'_{x,y,z} = \frac{n_{vx,vy,vz} \times s_{vx,vy,vz} - s_{vx,vy,vz}}{2}, \\
\mathbf x' = \text{Norm}((x,y,z)) = (-n'_x+(x-1) \times s_{vx},-n'_y+(y-1) \times s_{vy},-n'_z+(z-1) \times s_{vz}).
\end{gathered}
\end{equation}

\subsubsection{Multiresolution Hash Encoding}
\begin{figure}[!t] 
\centering
\includegraphics[width=\textwidth]{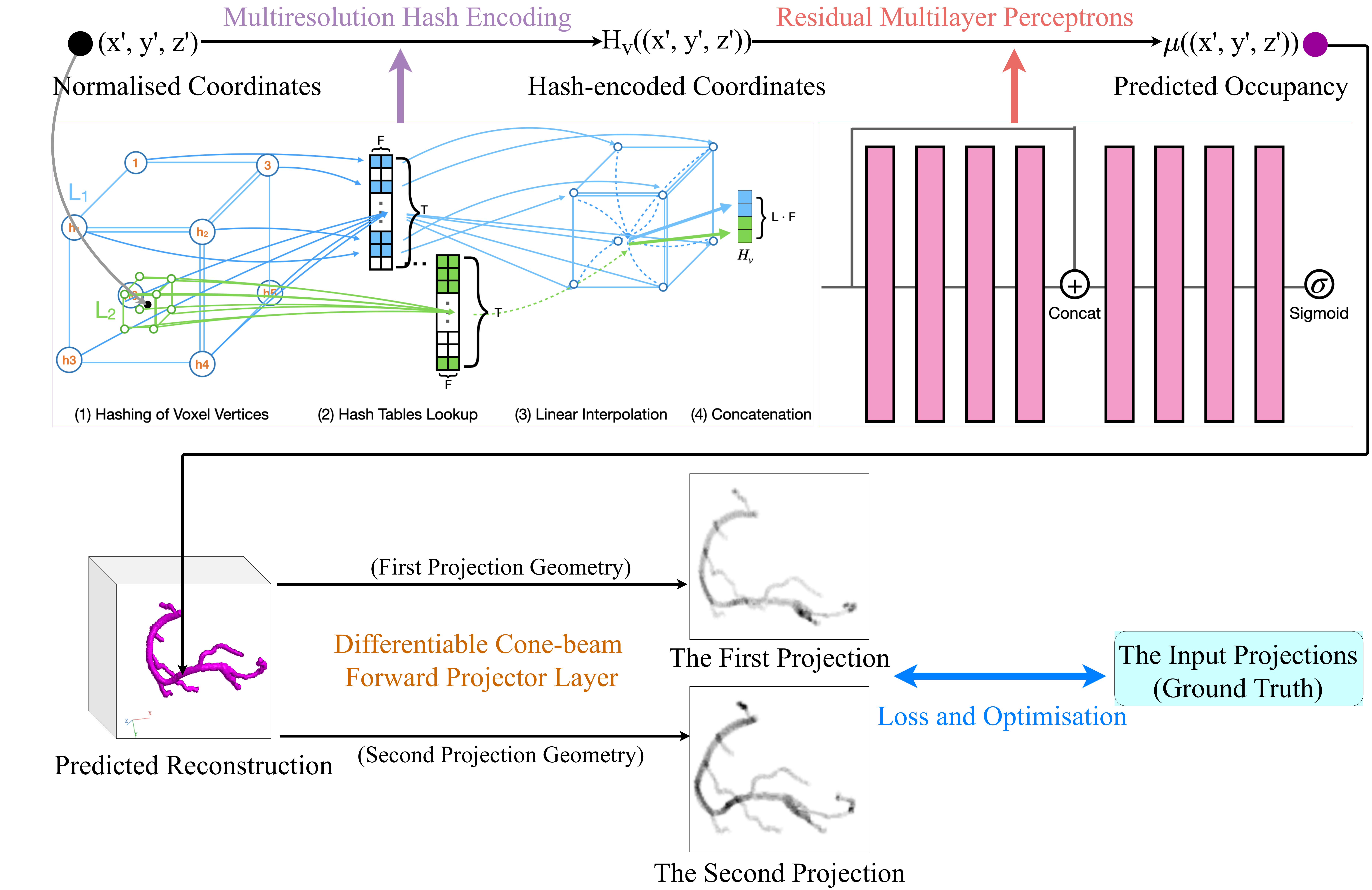}
\caption{The proposed NeCA model (stages 2--5). The multiresolution hash encoder illustrates an example of $2$ resolution levels (coloured in green and blue) from fine to coarse resolution for one sampled point (in black).} \label{model}
\end{figure}

We use the multiresolution hash encoder \cite{muller2022instant} $H_v = enc(\mathbf x'; \mathbf{\Theta})$ to encode the normalized positions of sampled points, which enables fast encoding without sacrificing performance. With the multiresolution structure, it allows the encoder to disambiguate hash collisions. The multiple resolutions are arranged into $L$ levels with different $T$-dimensional learnable hash tables at each level containing feature vectors with size $F$. The hyperparameters of our multiresolution hash encoder are shown in Table~\ref{hashencoder}, and the structure of the encoder is illustrated in Figure~\ref{model}. 
\begin{table}[!h]
\caption{The hyperparameters for the multiresolution hash encoder used in our work.}\label{hashencoder}
\centering
\begin{tabular}{lcc}
\toprule
\multicolumn{1}{c}{\textbf{Parameter}}                                              & \textbf{Symbol} & \textbf{Value} \\ \midrule
Number of levels                                                                      & $L$             & $16$           \\ \midrule
\begin{tabular}[c]{@{}l@{}}Maximum entries per level\\ (hash table size)\end{tabular} & $T$             & $2^{19}$       \\ \midrule
Number of feature dimensions per entry                                                & $F$             & $2$            \\ \midrule
Coarsest resolution                                                                   & $N_{min}$       & $16$           \\ \midrule
Resolution growth factor                                                              & $b$             & $2$            \\ \midrule
Input dimension                                                              & $d$             & $3$            \\ \bottomrule
\end{tabular}
\end{table}

\par For each voxel, we apply $L$ resolution levels, which are independent of each other. The resolution size $N$ is chosen based on an exponential increment between the coarsest and finest resolutions $\left \lfloor{N_{min},N_{max}}\right \rfloor$, where $N_{max}$ is selected to match the finest detail in the training data. It is defined as:
\begin{equation}
N_l \coloneqq \left \lfloor{N_{min}*b^l}\right \rfloor,
\end{equation}
where $l \in \{0,1, \dots L-1\}$, and $b = 2$ is the growth factor. For a single level $N_l$, the input point with normalized coordinates $\mathbf x' = (x',y',z') \in \mathbb{R}^3$ is geometrically scaled to a grid cube containing $2^3$ vertices according to the grid resolution at this level. To implement this functionality, the original 3D volume is evenly split into a number of grid cubes according to the resolution $N_l^3$, and the grid cube containing the desired sampled point is assigned to this point as the spanned grid cube. The multiresolution property in the hash encoder covers the full range from the coarsest resolution $N_{min}$ to the finest resolution $N_{max}$, which ensures that all scales are contained, in spite of sparsity. The four parts of the multiresolution hash encoder are discussed in detail below.

\paragraph{Hashing of Voxel Vertices} 
For all normalized voxels after scaling at resolution level $N_l$, we have $(N_l + 1)^d$ vertices in total. For coarse levels when $(N_l + 1)^d <= T$, we have one-to-one mapping from all the vertices at this resolution level $N_l$ to hash table entries, so there is no collision. Regarding finer levels when $(N_l + 1)^d > T$, we use a hash function $h$ to index into the feature vector array, effectively treating it as a hash table. In this case, we do not explicitly tackle hash collisions, but instead we reply on gradient-based optimization in the backpropagation of the subsequent residual MLP to automatically handle them. For instance, if two voxels have the same hash value on one or more vertices, the voxel closer to the desired object which our model is more focused on tends to have larger gradients during optimization, so this voxel takes the domination to update the collided feature vector entry. In this way, the collision issue is handled implicitly. 

\par We assign indices to these vertices by hashing their coordinates. The spatial hash function \cite{teschner2003optimized} $h$ is defined in the following form:
\begin{equation}
h(\mathbf x') = (\oplus_{i=1,2,3} \mathbf x'_i \pi_i)\mod T
\end{equation}
where $\mathbf x'$ is the input point, $\mathbf x'_{i=1,2,3}$ are the corresponding spatial normalized coordinate values, $\oplus$ denotes the bit-wise XOR operation, $\pi_i$ are unique large primary numbers, and $T$ is the hash table size. 

\paragraph{Hash Tables Lookup} 
We now have the hash value for each vertex at each resolution level of each point. We then maintain an individual learnable hash table, which contains $T$ numbers of $F$-dimensional feature vectors for each resolution level. For the hash values on all the vertices of each resolution level, we look up the corresponding entries in the level's respective feature vector array, i.e., the hash table. Next, the previously assigned indices on the vertices are replaced by the corresponding lookup feature vectors, so each resolution level conceptually stores feature vectors at the vertices of a grid cube. The hash tables at different resolution levels are the only trainable parameters $\mathbf{\Theta}$ in the multiresolution hash encoder, and the size of these parameters is $L \times T \times F$.

\paragraph{Linear Interpolation} 
For each resolution level, we linearly interpolate the feature vectors on the vertices according to their relative positions to the sampled point within this resolution level cube. Interpolating the queried hash table entries guarantees the encoded feature vectors with the later residual MLP are continuous during network training. After interpolation, the final feature vectors with the dimension $F$ for the sampled voxel at this resolution level are produced. 

\paragraph{Concatenation}
We concatenate the interpolated feature vectors for each resolution level to generate the final multiresolution hash encoding feature vectors $H_v \in \mathbb{R}^{L \times F}$ for the sampled point, which can then be utilised to predict the occupancy of coronary tree for this point position by the residual MLP at the next stage. The dimension $L \cdot F$ for the final encoded feature vectors of each voxel is regarded as the channel dimension for later residual MLP training.

\subsubsection{Residual MLP} 
We exploit residual \cite{he2016deep} MLP $m(H_v; \mathbf{\Phi})$ to predict the occupancy value $\mu$ from the position-encoded feature vectors $H_v$ of each point, where $\mathbf{\Phi}$ is the trainable weight parameters of the residual MLP. The residual MLP network serves as a continuous function to implicitly parameterise a bounded scene, i.e., the 3D coronary tree in our case, which maps spatial coordinate features to the predicted occupancy values. This, in fact, encodes the internal information of an entire 3D coronary tree into the network parameters.

\par The residual MLP contains eight fully connected layers, as depicted in Figure~\ref{model}. We apply residual learning in the middle layer to preserve the original feature information. The residual MLP receives the feature vectors as input with $L \cdot F$-dimensional channels and produces predicted occupancy values with a $1$-dimensional channel. The feature dimensions for all the hidden layers are $256$-wide. Except for the last layer followed by a sigmoid activation, all the layer outputs are followed by LeakyReLU activation \cite{maas2013rectifier}.

\subsubsection{Differentiable Forward Projector Layer}
At this stage, we have all the predicted occupancy values for all the voxels, which construct the 3D coronary tree reconstruction results. After that, we simulate the X-ray cone-beam forward projections from the 3D reconstruction results based on the same projection geometry as the input projections to generate two predicted projections. The forward projection simulation is based on the theory that the intensity of an X-ray beam is reduced by the exponential integration of attenuation coefficients along the ray path. We use ODL \cite{jonas_adler_2017_249479} to implement this differentiable X-ray forward projector layer that enables self-supervised loss optimization at the final stage.

\subsubsection{Loss}
We use Mean Square Error (MSE) loss to calculate the differences between the input projections and simulated forward projections. The loss function $\mathcal L$ is defined as follows:
\begin{equation}
\mathcal{L}(\mathbf{\Theta}, \mathbf{\Phi}) = \frac{1}{N \times I}\sum^N_n \sum^I_i (P_{ni} - G_{ni})^2
\end{equation}
where $N$ ($= 2$ in our work) is the number of projections, $I$ ($= 512$ in our work) is the number of pixels in one projection, $P$ is the simulated projection, and $G$ is the corresponding input projection.

\par The loss function is used to learn the multiresolution hash tables $\mathbf{\Theta}$ and the residual MLP $\mathbf{\Phi}$ during training. With this, the 3D occupancy predictions are improved iteratively based on the optimization of 2D projection errors. After training, the final 3D coronary tree can be rendered with the predicted occupancy values, after binarisation with $0.5$, by querying all the voxels with their coordinates from the model.

\subsection{Training Setup}
We implement our proposed model using PyTorch \cite{paszke2019pytorch} and choose the Adam optimiser \cite{kingma2014adam} with a learning rate of $10^{-4}$. The number of epochs for optimization is 5000. The learning was performed on an HPC cluster utilizing Nvidia Tesla v100 GPUs. {The package versions we used for NeCA are Python 3.8.17, PyTorch 1.9.0, and ODL 1.0.0.dev0.}

\subsection{Baseline Model}
We use the supervised learning model 3D U-Net \cite{cciccek20163d} as our baseline model. We follow the original 3D U-Net architecture with three sampling levels and a bottleneck layer using the same number of convolutional filters. The channel size for both the input and output to 3D U-Net model in our work is 1. The input to 3D U-Net is an ill-posed volume reconstructed from two clinical-angle projections of the 3D coronary tree by a conventional back-projection method, and the output is the 3D coronary tree reconstruction result. We train two 3D U-Net models based on the CCTA dataset \cite{ZENG2023102287} using 669 RCA data points and 788 LAD data points, respectively, where we split them into 75\% training, 15\% validation, and 10\% test data. The test datasets here are the same datasets used for testing our proposed model.

\par We implement the 3D U-Net baseline model using PyTorch \cite{paszke2019pytorch} and choose the Adam optimiser \cite{kingma2014adam} with an initial learning rate of $10^{-4}$. A learning rate decay policy is used, where the learning rate is decayed by $0.1$ if no improvement is observed after 10 epochs. We use an early stopping strategy to avoid overfitting when there is no more improvement after 15 epochs. The training was performed with a batch size of 3 on an HPC cluster utilizing Nvidia Tesla v100 GPUs. The models are trained with MSE loss.

\subsection{Evaluation Metrics}
We employ six metrics for evaluation between the 3D coronary tree reconstruction results and the original CCTA data (ground truth): centerline Dice score (termed as \emph{clDice}) \cite{shit2021cldice}, Dice score (termed as \emph{Dice}), intersection over union (termed as \emph{IoU}), reconstruction error (termed as \emph{reError})~\cite{bousse2008motion}, Chamfer $\ell_2$ distance (termed as \emph{CD$_{\ell_2}$}), and reconstruction MSE (termed as \emph{reMSE}). $\text{\emph{clDice}} \in [0,1]$ where a larger value suggests a better performance in vessel topology preservation. \emph{Dice} ($\in [0,1]$) and \emph{IoU} ($\in [0,1]$) also suggest a better performance if measurement values are bigger. In terms of \emph{reError}, \emph{CD$_{\ell_2}$}, and \emph{reMSE}, a smaller value represents a better reconstruction result. Before evaluation, we apply connected component analysis \cite{silversmith2021cc3d} on our reconstructed coronary tree to remove sparse disconnected objects with less than $25$ voxels.

\section{Results}
We perform both quantitative and qualitative evaluations on both RCA and LAD datasets. Apart from the clinical-angle projections simulated according to Table~\ref{proj_geo}, we additionally test 3D reconstructions based on two orthogonal views using our NeCA model for comparison (termed as NeCA (90$\degree$)). 

\subsection{Quantitative Results}
We quantitatively evaluate our NeCA model, NeCA (90$\degree$), and supervised 3D U-Net model on 67 RCA test data points and 79 LAD test data points.

\subsubsection{RCA Dataset}
\label{analysis_rca}
\paragraph{Performance over Six Metrics}
\label{metrics_eval_rca}
We evaluate NeCA, NeCA (90$\degree$), and the 3D supervised U-Net model in terms of six metrics, namely \emph{clDice}, \emph{Dice}, \emph{IoU}, \emph{reError}, \emph{CD$_{\ell_2}$}, and \emph{reMSE}. The quantitative results are presented in Table~\ref{rca_numerical}.

\begin{table}[!h]
\caption{The quantitative evaluation results of NeCA, NeCA (90$\degree$), and supervised 3D U-Net model on 67 RCA test data in terms of six metrics. The best results of each metric are in \textbf{bold}.}\label{rca_numerical}
\centering 
\begin{adjustbox}{width=\linewidth}
\begin{tabular}{lcccccc}
\toprule
\textbf{Model}           & \emph{\textbf{clDice}} (\%)  & \emph{\textbf{Dice}} (\%)  &\emph{\textbf{IoU}} (\%)  & \emph{\textbf{reError}}                                        & \emph{\textbf{CD}\boldmath{$_{\ell_2}$}} ($mm$) & \emph{\textbf{reMSE}} ($1 \times 10^{-4}$) \\ \midrule
NeCA   &87.01 $\pm$ 9.93 
        & 90.43 $\pm$ 7.46              &83.29 $\pm$ 11.42           & 0.139 $\pm$ 0.101        & 0.27 $\pm$ 0.37               & 2.74 $\pm$ 2.14 \\
NeCA (90$\degree$) & 89.07 $\pm$ 8.33             &\textbf{91.03} $\pm$ 6.93     &\textbf{84.17} $\pm$ 10.25  &\textbf{0.111} $\pm$ 0.087 &\textbf{0.22} $\pm$ 0.26                          & \textbf{2.73} $\pm$ 2.60                           \\
3D U-Net           & \textbf{95.34} $\pm$ 4.16      & 85.18 $\pm$ 4.22 &74.42 $\pm$ 6.24     & 0.188 $\pm$ 0.054      & 0.31 $\pm$ 0.16        & 4.63 $\pm$ 2.91                                   \\ \bottomrule
\multicolumn{7}{@{}l}{All values represent mean ($\pm$ standard deviation).}
\end{tabular}
\end{adjustbox}
\end{table} 

\par From the results presented in Table~\ref{rca_numerical}, we can observe that our NeCA model performs better than 3D U-Net model, with relative improvements of 6.16\%, 11.92\%, 26.06\%, 12.90\%, and 40.82\% in terms of \emph{Dice}, \emph{IoU}, \emph{reError}, \emph{CD$_{\ell_2}$}, and \emph{reMSE} metrics, respectively. 3D U-Net model is better than our NeCA model based on the \emph{clDice} metric, with a respective improvement of 9.57\%. 3D reconstruction from two orthogonal projections by our NeCA model produces the best performance in all metrics compared to our NeCA model using two clinical-angle projections. 3D U-Net model maintains the smallest standard deviations among all metrics except for \emph{reMSE}, where our NeCA model performs the best.

\paragraph{Statistical Analysis}
\label{aso_rca_section}
The choice of the statistical test is very important, as different tests can have different conclusions for the same evaluation. For this reason and the nature of deep learning in our work, we use the Almost Stochastic Order (ASO) test \cite{del2018optimal, dror2019deep} as implemented by \cite{ulmer2022deep} specifically for deep leaning models to compare score distributions from different models, with a significance level $\alpha$. ASO returns a confidence score $\epsilon_\text{min}$, which indicates (an upper bound to) the amount of violation of stochastic order. In terms of analysis between model A and B using ASO, if $\epsilon_\text{min} < \tau$ (where the rejection threshold $\tau$ is $0.5$ or less), model A is said to be stochastically dominant over model B in more cases, and model A is considered superior. The lower $\epsilon_\text{min}$ is, the more confidently we can conclude that model A outperforms model B. The tests from \cite{ulmer2022deep} show that $\tau = 0.2$ is the most effective threshold value that has a satisfactory tradeoff between Type I and Type II errors across different scenarios. Please note for metrics such as errors where a smaller value expresses a better performance, the final confidence score $\epsilon_\text{min}$ should be $1$ minus the returned $\epsilon_\text{min}$ from ASO.

\par With regard to statistical significance test in our work using ASO, we choose a significance level $\alpha = 0.05$ and $\tau = 0.2$. The confidence scores for all six metrics between our NeCA model and 3D U-Net model using the ASO testing on the RCA test dataset are demonstrated in Table~\ref{aso_rca}.
\begin{table}[!h]
\caption{The confidence scores $\epsilon_\text{min}$ for six metrics between our NeCA model and 3D U-Net model using the ASO testing with a significance level $\alpha = 0.05$ on the RCA test dataset. The confidence scores where our NeCA model is found to be stochastically dominant over 3D U-Net are in \textbf{bold}, i.e., $\epsilon_\text{min} < \tau = 0.2$.}\label{aso_rca}
\centering
\begin{tabular}{cccccccc}
\toprule
                      & \emph{\textbf{clDice}} & \emph{\textbf{Dice}} & \emph{\textbf{IoU}} & \emph{\textbf{reError}} & \emph{\textbf{CD}\boldmath{$_{\ell_2}$}} & \emph{\textbf{reMSE}} \\ \midrule
$\epsilon_\text{min}$ & 0.982350                           & \textbf{0.198873}                         & \textbf{0.127973}                        & \textbf{0.0}                               & 0.287172                                  & \textbf{0}                             \\
\bottomrule
\end{tabular}
\end{table}

\par From Table~\ref{aso_rca}, we can find that the score distributions of our NeCA model in terms of \emph{Dice}, \emph{IoU}, \emph{reError}, and \emph{reMSE} are stochastically dominant over the 3D U-Net model. Regarding the metric \emph{CD$_{\ell_2}$}, according to threshold $\tau = 0.2$, our NeCA model is better but not stochastically dominant over 3D U-Net. For \emph{clDice}, the 3D U-Net model is found to be stochastically dominant over our NeCA model.

\begin{figure}[!h]
	\begin{subfigure}[h]{0.48\textwidth}
         \centering
        \includegraphics[width=\textwidth]{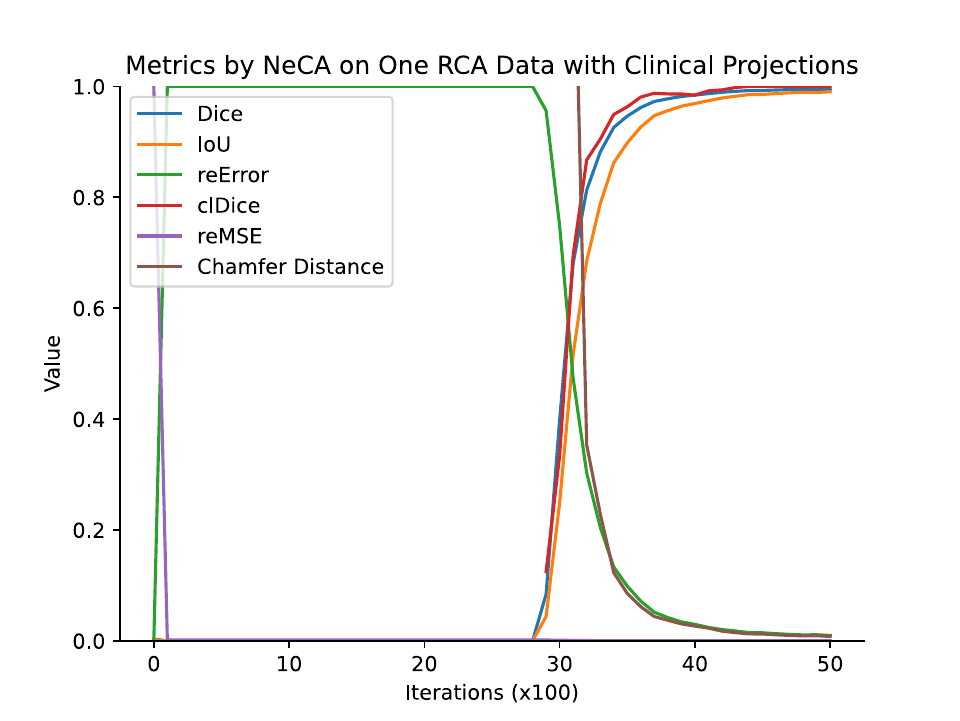}
     \end{subfigure}
     \hfill
	\begin{subfigure}[h]{0.48\textwidth}
         \centering
    \includegraphics[width=\textwidth]{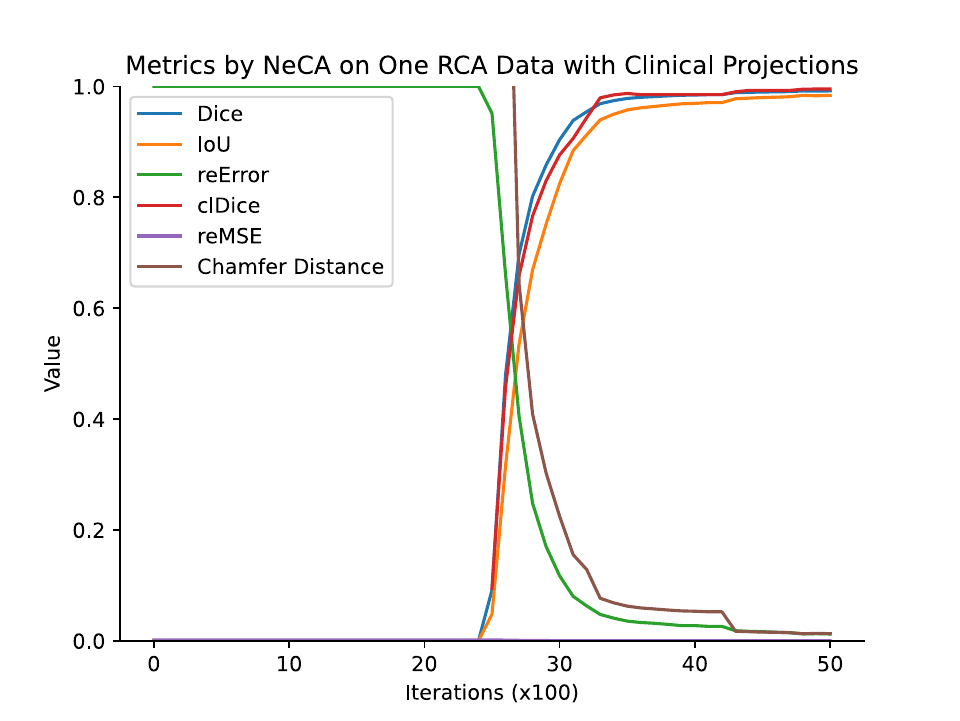}
     \end{subfigure}
    \caption{The results of all six metrics every 100 iterations for two RCA example data points ($R_1$ and $R_2$) using our NeCA model with two clinical-angle projections.}
        \label{iter_results_rca_clinical}
\end{figure}

\paragraph{Optimizing the Performance of our NeCA Model over Iterations}
\label{rca_iter_performance}
Our NeCA model is optimised for each individual data point, and we record the quantitative evaluation results of different metrics every 100 iterations. Here, we use two RCA example data points to show how the performance improves iteratively using our NeCA model with clinical-angle projections, as illustrated in Figure~\ref{iter_results_rca_clinical}.

\par We can see from Figure~\ref{iter_results_rca_clinical} that the performance starts to improve after 2000 iterations. We can also find that it usually takes less than 2000 iterations to reach good results after the improvement starts.

\subsubsection{LAD Dataset}
\paragraph{Performance over Six Metrics}
We perform the quantitative evaluations on the LAD test dataset the same as for the RCA data, as described in Section~\ref{analysis_rca}. The results are presented in Table~\ref{lad_numerical}.
\begin{table}[!h]
\caption{The quantitative evaluation results of NeCA, NeCA (90$\degree$), and 3D U-Net model on 79 LAD test data points in terms of 6 metrics. The best results of each metric are in \textbf{bold}.}\label{lad_numerical}
\centering 
\begin{adjustbox}{width=\linewidth}
\begin{tabular}{lcccccc}
\toprule
\textbf{Model}     & \emph{\textbf{clDice}} (\%) & \emph{\textbf{Dice}} (\%) & \emph{\textbf{IoU}} (\%) & \emph{\textbf{reError}}                                         & \emph{\textbf{CD}\boldmath{$_{\ell_2}$}} ($mm$) & \emph{\textbf{reMSE}} ($1 \times 10^{-4}$) \\ \midrule
NeCA               & 76.08 $\pm$ 10.42 
                & 77.48 $\pm$ 9.93              & 64.28 $\pm$ 13.00            & 0.322 $\pm$ 0.129         & 0.75 $\pm$ 0.49                          & 7.28 $\pm$ 3.61                                         \\
NeCA (90$\degree$) & \textbf{91.69}$\pm$ 5.62        & \textbf{94.27} $\pm$ 3.91      & \textbf{89.41}$\pm$ 6.70     & \textbf{0.077} $\pm$ 0.051 & \textbf{0.17} $\pm$ 0.18                    & \textbf{2.26} $\pm$ 1.89                              \\
3D U-Net           & 83.36 $\pm$ 7.50                & 68.54 $\pm$ 6.87              & 52.54 $\pm$ 7.91             & 0.415 $\pm$ 0.081         & 0.99 $\pm$ 0.51                            & 10.38 $\pm$ 4.22                                        \\ \bottomrule
\multicolumn{7}{@{}l}{All values represent mean ($\pm$ standard deviation).}
\end{tabular}
\end{adjustbox}
\end{table}

\par In Table~\ref{lad_numerical}, in contrast to the 3D U-Net model, our NeCA model shows improvements of 13.04\%, 22.34\%, 22.41\%, 24.24\%, and 29.87\% in terms of \emph{Dice}, \emph{IoU}, \emph{reError}, \emph{CD$_{\ell_2}$}, and \emph{reMSE}, respectively. The 3D U-Net model is 9.57\% better than our NeCA model with respect to \emph{clDice}. Our NeCA model with two orthogonal projections as input maintains the best performance among all six metrics compared to both our NeCA model with clinical-angle projections and the 3D U-Net model. Furthermore, our NeCA model with two orthogonal projections as input has the smallest standard deviations among all six metrics compared to both the 3D U-Net model and NeCA with clinical-angle projections.

\paragraph{Statistical Analysis}
For the statistical significance analysis on the LAD test dataset, we use the ASO test, as described in Section~\ref{analysis_rca}, where we choose a significance level of $\alpha = 0.05$ and $\tau = 0.2$. The confidence scores in terms of all six metrics between our NeCA model and the 3D U-Net model are presented in Table~\ref{aso_lad}.
\begin{table}[!h]
\caption{The confidence scores $\epsilon_\text{min}$ for six metrics between our NeCA model and the 3D U-Net model on the LAD test dataset using ASO testing with a significance level of $\alpha = 0.05$. The confidence scores where our NeCA model is tested to be stochastically dominant over 3D U-Net are in \textbf{bold}, i.e., $\epsilon_\text{min} < \tau = 0.2$.}\label{aso_lad}
\centering
\begin{tabular}{ccccccc}
\toprule
                      & \emph{\textbf{clDice}} & \emph{\textbf{Dice}} & \emph{\textbf{IoU}} & \emph{\textbf{reError}} & \emph{\textbf{CD}\boldmath{$_{\ell_2}$}} & \emph{\textbf{reMSE}} \\ \midrule
$\epsilon_\text{min}$ & 0.992092                           & \textbf{0.010340}                         & \textbf{0.005389}                        & \textbf{0}                               & \textbf{0}                                 & \textbf{0}                             \\
\bottomrule
\end{tabular}
\end{table}

\par Table~\ref{aso_lad} demonstrates that our NeCA model evidently outperforms the 3D U-Net model in terms of five metrics, namely \emph{Dice}, \emph{IoU}, \emph{reError}, \emph{CD$_{\ell_2}$}, and \emph{reMSE}. In terms of the \emph{clDice} metric, the 3D U-Net model is stochastically dominant over the NeCA model.

\begin{figure}[!h]
	\begin{subfigure}[h]{0.48\textwidth}
         \centering
\includegraphics[width=\textwidth]{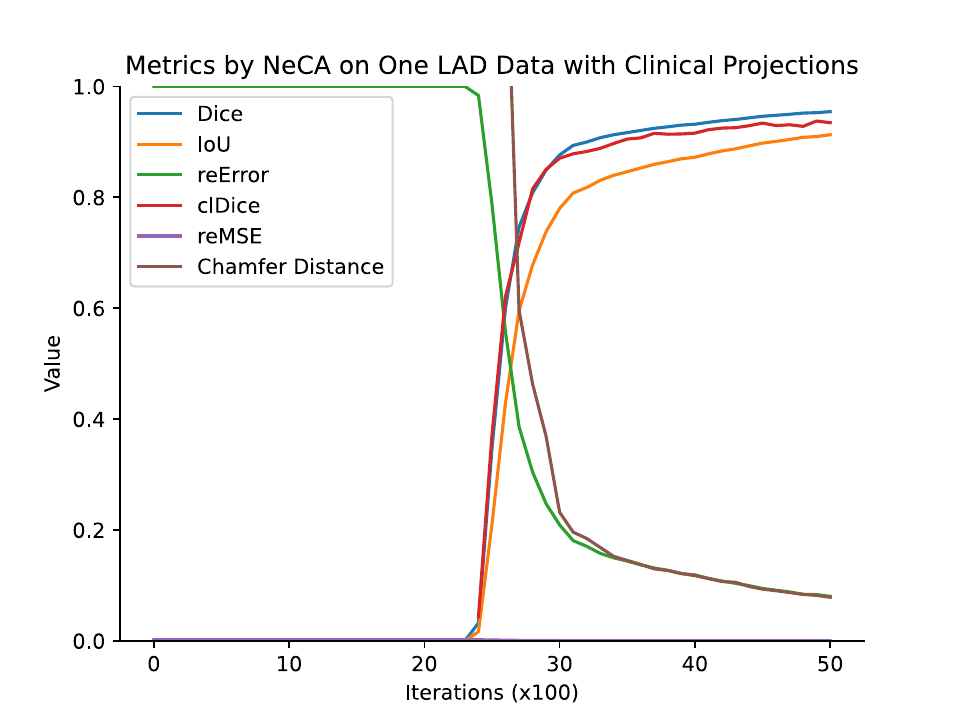}
     \end{subfigure}
     \hfill
	\begin{subfigure}[h]{0.48\textwidth}
         \centering
\includegraphics[width=\textwidth]{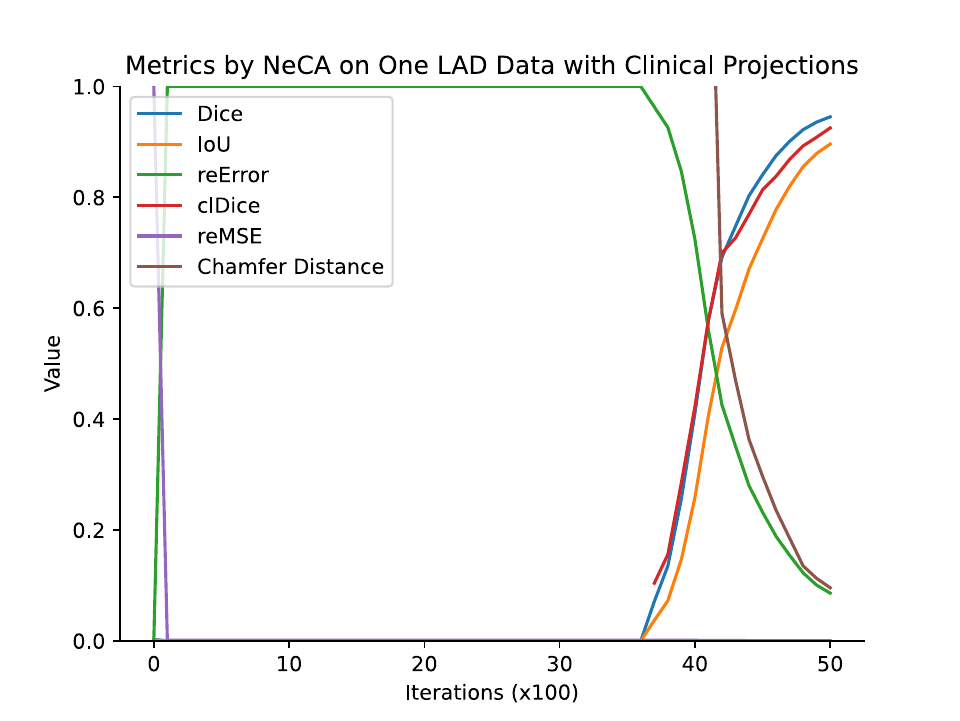}
     \end{subfigure}
    \caption{The quantitative results of our NeCA model over two LAD example data points ($L_2$ and $L_3$) every 100 iterations with respect to all 6 metrics and evaluated with 2 clinical-angle projections.}
        \label{iter_results_lad_clinical}
\end{figure}

\paragraph{Optimizing the Performance of our NeCA Model Over Iterations}
We record the quantitative evaluation results of different metrics every 100 iterations for each individual data point our NeCA model optimises for. Here, we report two LAD example data points to demonstrate how the NeCA model's performance improves iteratively, as illustrated in Figure~\ref{iter_results_lad_clinical}.

\par From Figure~\ref{iter_results_lad_clinical}, we can see that the performance does not start to improve until at least 2000 iterations, and it often takes about 2000 iterations to reach satisfactory performance after the improvement starts. The same phenomenon is also observed for the RCA dataset in Section~\ref{analysis_rca}.

\subsection{Qualitative Results}\label{qualitative_results}
We present the qualitative results of 3D coronary artery tree reconstruction based on our NeCA model, NeCA (90$\degree$), and the 3D U-Net model on both the RCA and LAD test datasets. Here, we use five example data points for each dataset. 

\subsubsection{RCA Dataset}
\begin{figure}[!h]
     \hspace{0.3cm}NeCA\hspace{0.3cm}
     \begin{subfigure}[h]{0.17\textwidth}
         \centering
         \includegraphics[width=\textwidth]{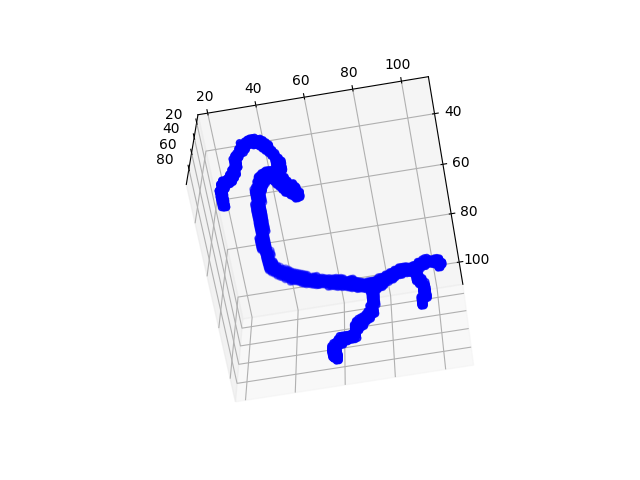}
     \end{subfigure}
     \hfill
     \begin{subfigure}[h]{0.17\textwidth}
         \centering
         \includegraphics[width=\textwidth]{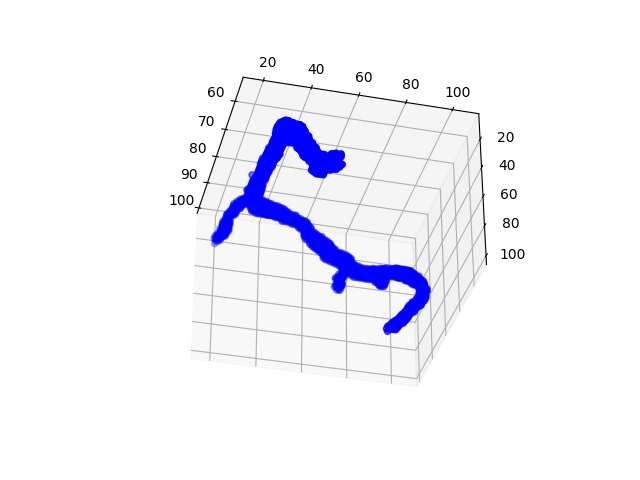}
     \end{subfigure}
     \hfill
     \begin{subfigure}[h]{0.17\textwidth}
         \centering
         \includegraphics[width=\textwidth]{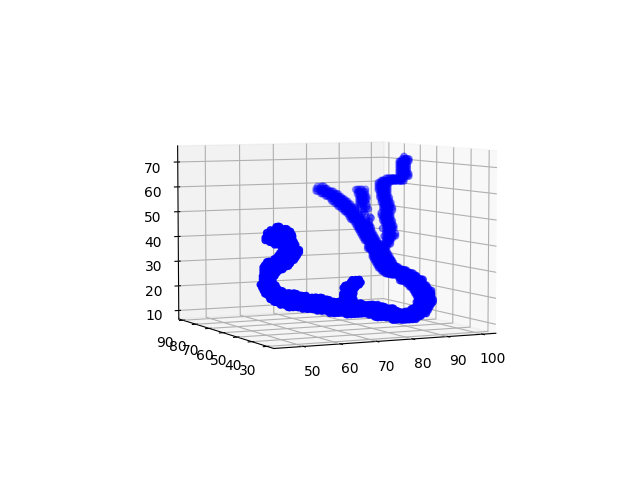}
     \end{subfigure}
     \hfill
     \begin{subfigure}[h]{0.17\textwidth}
         \centering
         \includegraphics[width=\textwidth]{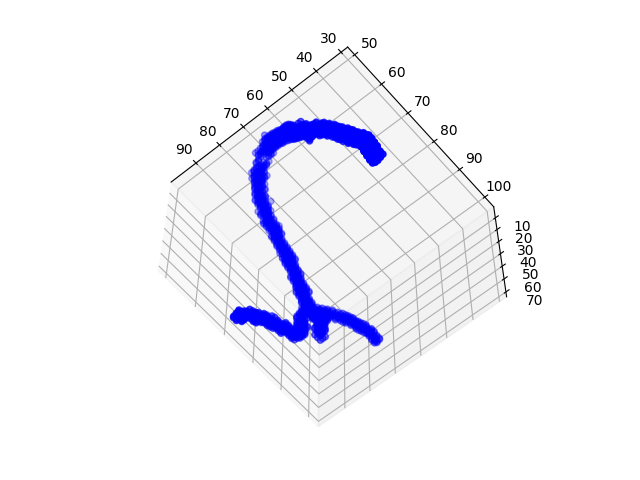}
     \end{subfigure}
     \hfill
     \begin{subfigure}[h]{0.17\textwidth}
         \centering
         \includegraphics[width=\textwidth]{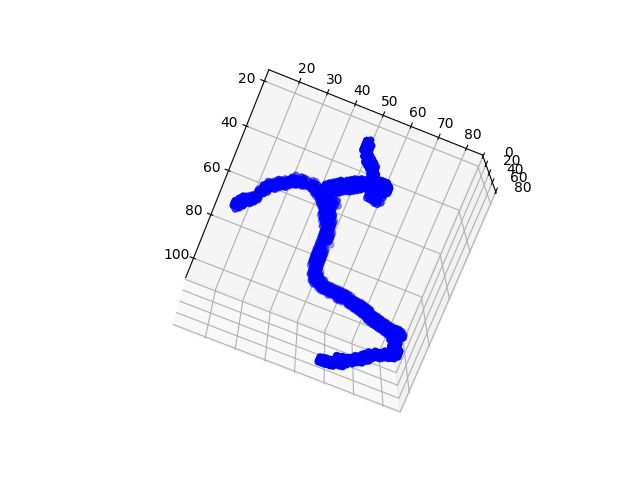}
     \end{subfigure}
     \vfill
     NeCA (90$\degree$)\hspace{-0.11cm}
     \begin{subfigure}[h]{0.17\textwidth}
         \centering
         \includegraphics[width=\textwidth]{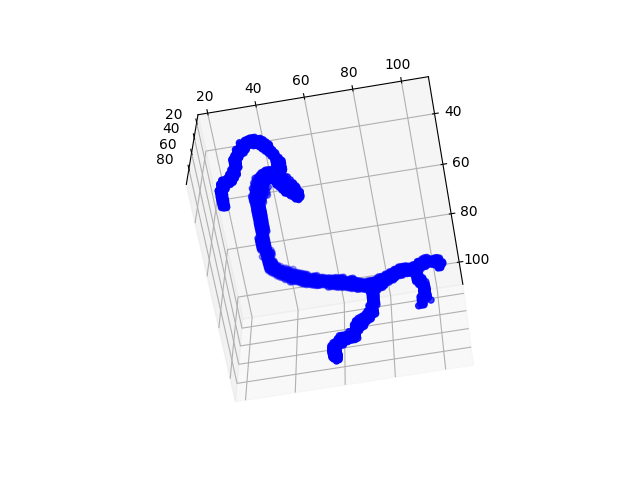}
     \end{subfigure}
     \hfill
	\begin{subfigure}[h]{0.17\textwidth}
         \centering
         \includegraphics[width=\textwidth]{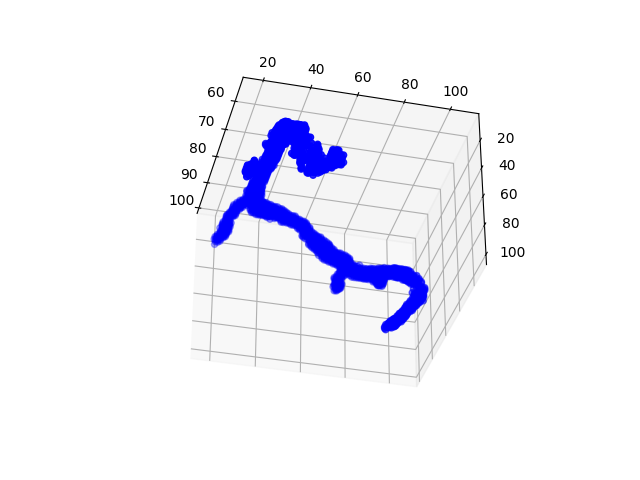}
     \end{subfigure}
     \hfill
	\begin{subfigure}[h]{0.17\textwidth}
         \centering
         \includegraphics[width=\textwidth]{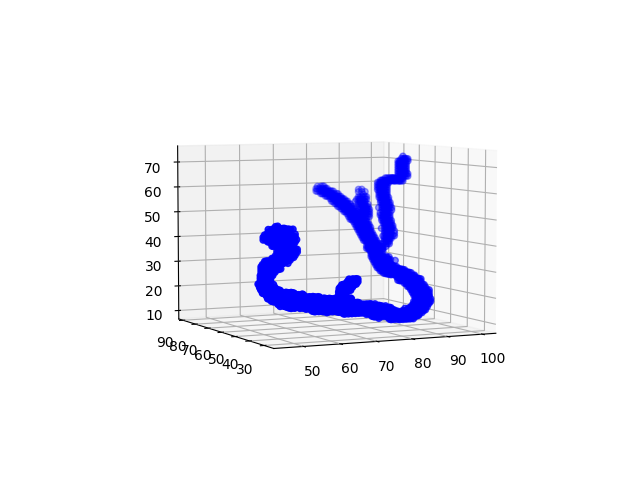}
     \end{subfigure}
     \hfill
	\begin{subfigure}[h]{0.17\textwidth}
         \centering
         \includegraphics[width=\textwidth]{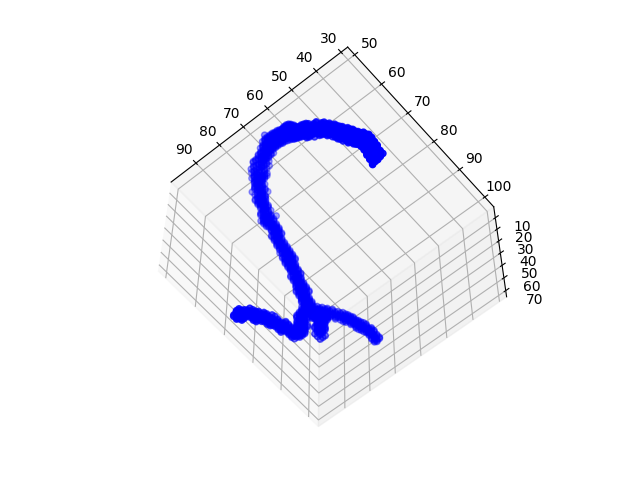}
     \end{subfigure}
     \hfill
	\begin{subfigure}[h]{0.17\textwidth}
         \centering
         \includegraphics[width=\textwidth]{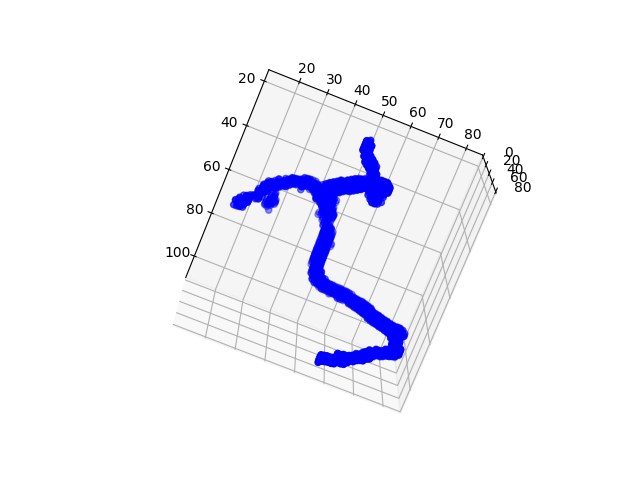}
     \end{subfigure}
     \vfill
     \hspace{0.05cm}3D U-Net\hspace{0.05cm}
	\begin{subfigure}[h]{0.17\textwidth}
         \centering
         \includegraphics[width=\textwidth]{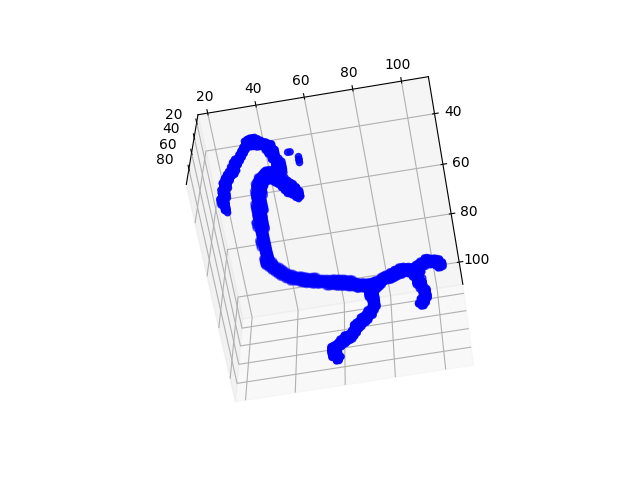}
     \end{subfigure}
     \hfill
     \begin{subfigure}[h]{0.17\textwidth}
         \centering
         \includegraphics[width=\textwidth]{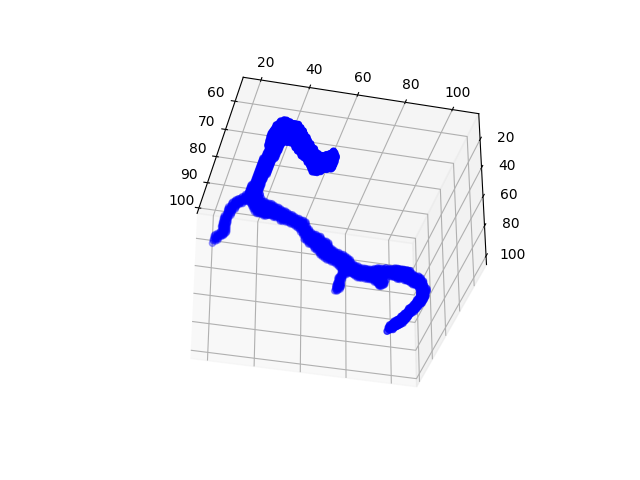}
     \end{subfigure}
     \hfill
	\begin{subfigure}[h]{0.17\textwidth}
         \centering
         \includegraphics[width=\textwidth]{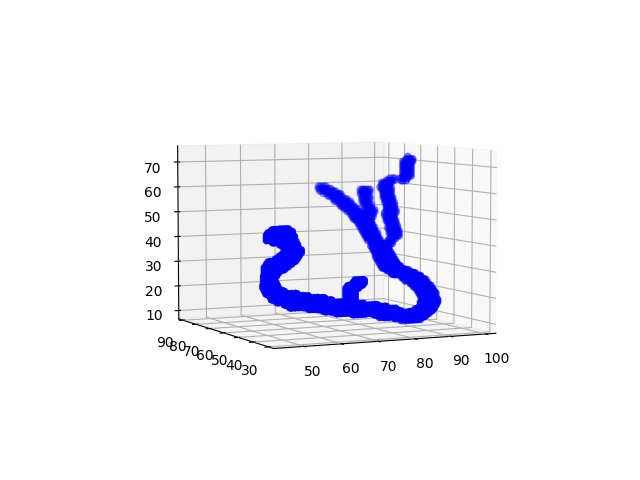}
     \end{subfigure}
     \hfill
	\begin{subfigure}[h]{0.17\textwidth}
         \centering
         \includegraphics[width=\textwidth]{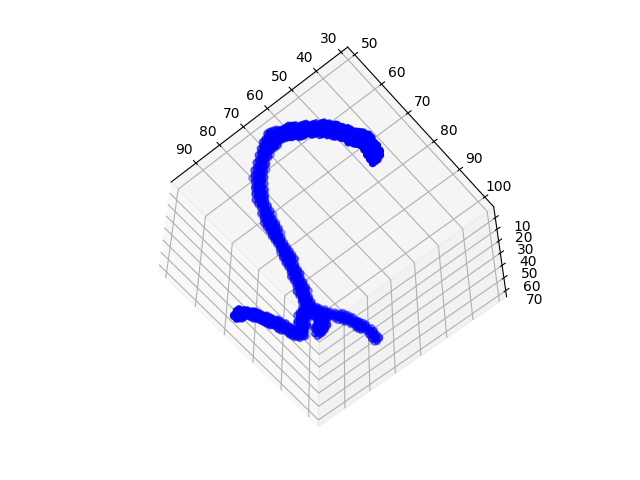}
     \end{subfigure}
     \hfill
	\begin{subfigure}[h]{0.17\textwidth}
         \centering
         \includegraphics[width=\textwidth]{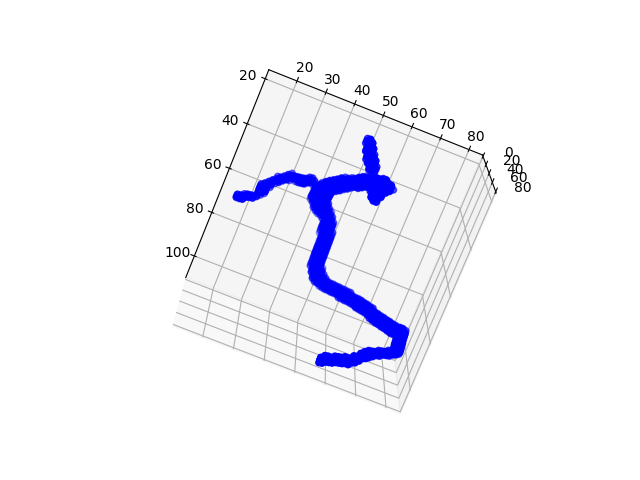}
     \end{subfigure}
     \vfill
     \hspace{0.55cm}GT\hspace{0.55cm}
	\begin{subfigure}[h]{0.17\textwidth}
         \centering
         \includegraphics[width=\textwidth]{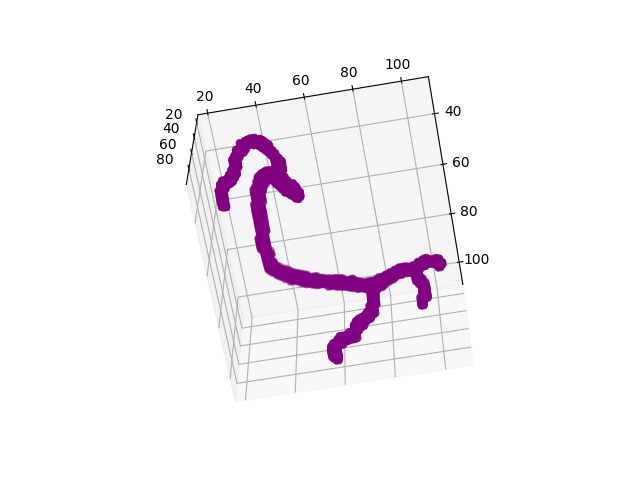}
         \caption*{\hspace{1cm}$R_1$}
     \end{subfigure}
     \hfill
	\begin{subfigure}[h]{0.17\textwidth}
         \centering
         \includegraphics[width=\textwidth]{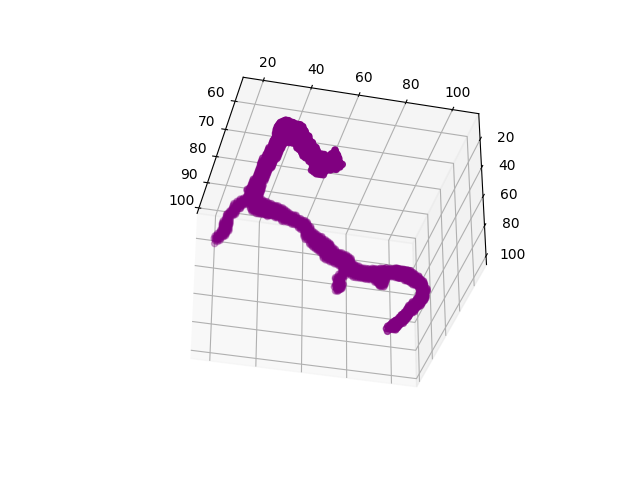}
         \caption*{\hspace{1cm}$R_2$}
     \end{subfigure}
     \hfill
	\begin{subfigure}[h]{0.17\textwidth}
         \centering
         \includegraphics[width=\textwidth]{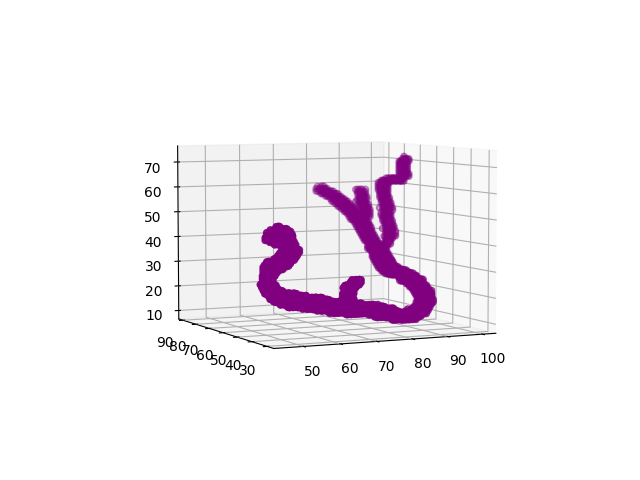}
         \caption*{\hspace{1cm}$R_3$}
     \end{subfigure}
     \hfill
	\begin{subfigure}[h]{0.17\textwidth}
         \centering
         \includegraphics[width=\textwidth]{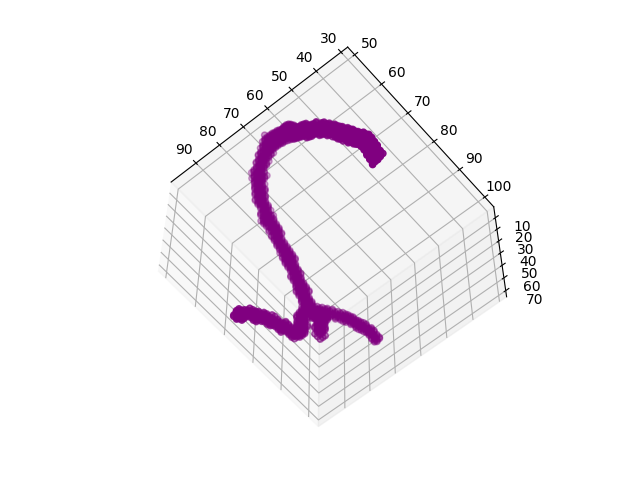}
         \caption*{\hspace{1cm}$R_4$}
     \end{subfigure}
     \hfill
	\begin{subfigure}[h]{0.17\textwidth}
         \centering
         \includegraphics[width=\textwidth]{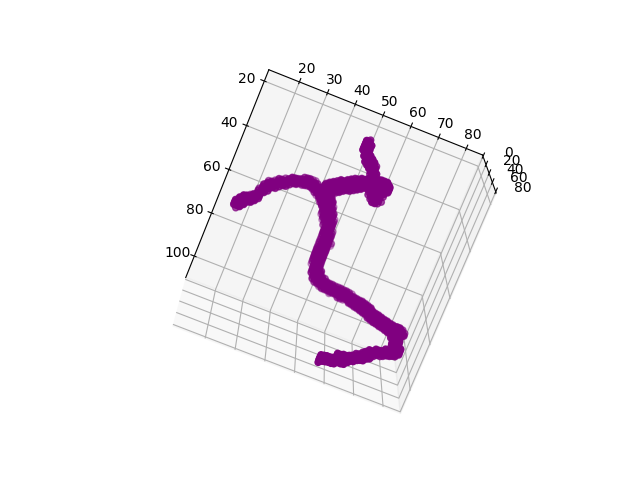}
         \caption*{\hspace{1cm}$R_5$}
     \end{subfigure}
    \caption{Five qualitative results of 3D RCA reconstruction. From left to right: five RCA data points $R_{1,2,3,4,5}$. From top to bottom: the reconstruction results from our NeCA model, NeCA ($90\degree$), 3D U-Net model, and the corresponding ground truth (GT).}
        \label{visual_recon_rca}
\end{figure}

\paragraph{3D Reconstruction Results}
Figure~\ref{visual_recon_rca} illustrates five RCA examples of 3D coronary tree reconstruction using our NeCA model, NeCA (90$\degree$), and 3D U-Net model, along with the corresponding ground truth for each case. The results show that all three models can successfully perform satisfactory 3D RCA reconstruction.

\begin{figure}[!h]
     \hspace{0.3cm}NeCA\hspace{0.3cm}
     \begin{subfigure}[h]{0.17\textwidth}
         \centering
         \includegraphics[width=\textwidth]{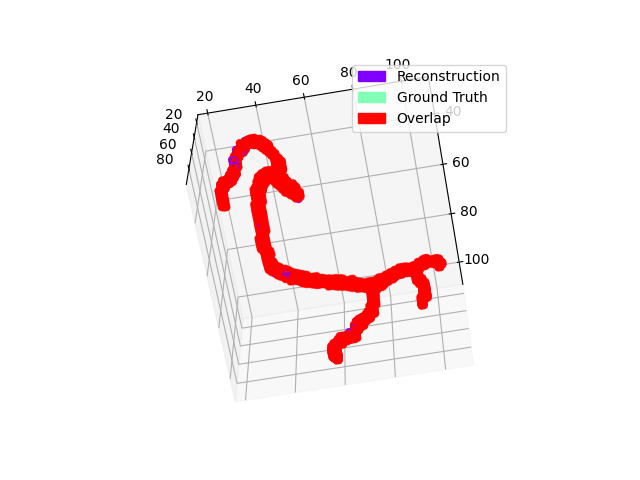}
     \end{subfigure}
     \hfill
	\begin{subfigure}[h]{0.17\textwidth}
         \centering
         \includegraphics[width=\textwidth]{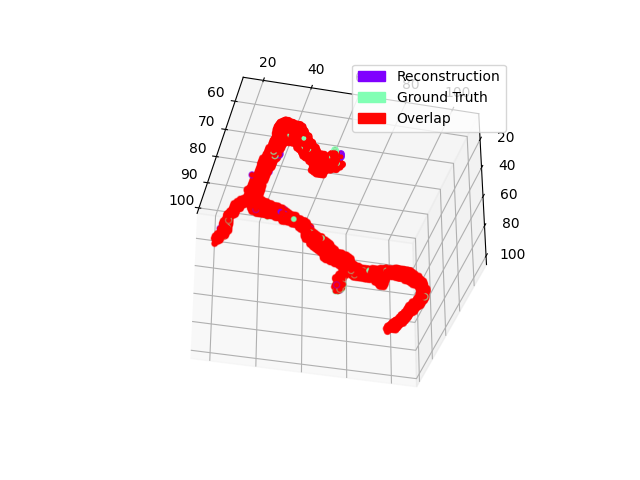}
     \end{subfigure}
     \hfill
	\begin{subfigure}[h]{0.17\textwidth}
         \centering
         \includegraphics[width=\textwidth]{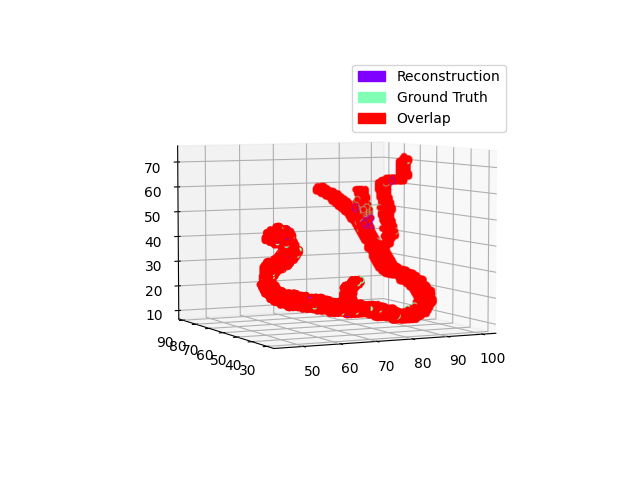}
     \end{subfigure}
     \hfill
	\begin{subfigure}[h]{0.17\textwidth}
         \centering
         \includegraphics[width=\textwidth]{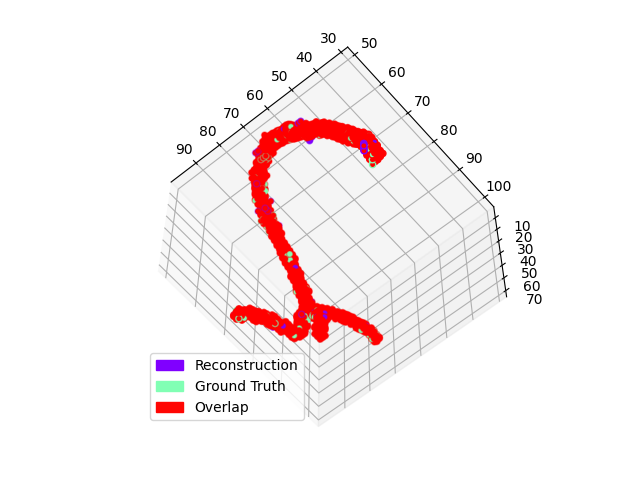}
     \end{subfigure}
     \hfill
	\begin{subfigure}[h]{0.17\textwidth}
         \centering
         \includegraphics[width=\textwidth]{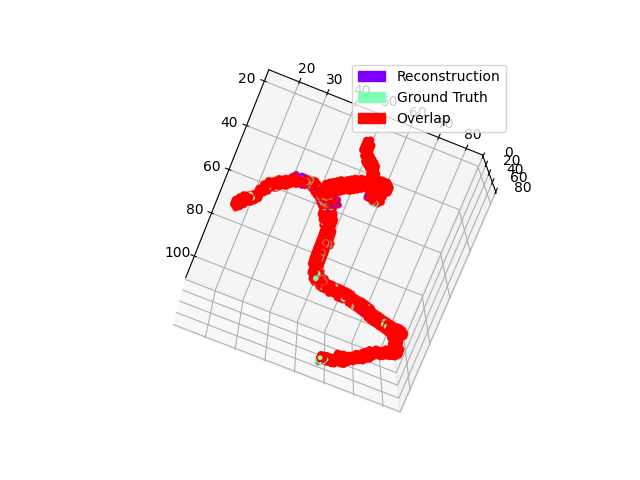}
     \end{subfigure}
     \vfill
	NeCA ($90\degree$)\hspace{-0.11cm}
	\begin{subfigure}[h]{0.17\textwidth}
         \centering
         \includegraphics[width=\textwidth]{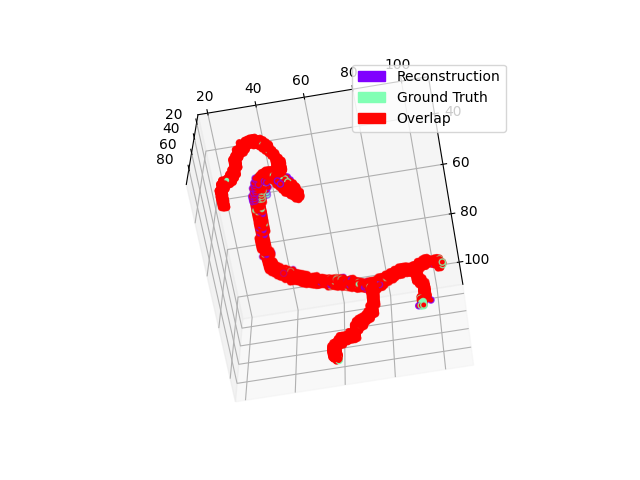}
     \end{subfigure}
     \hfill
	\begin{subfigure}[h]{0.17\textwidth}
         \centering
         \includegraphics[width=\textwidth]{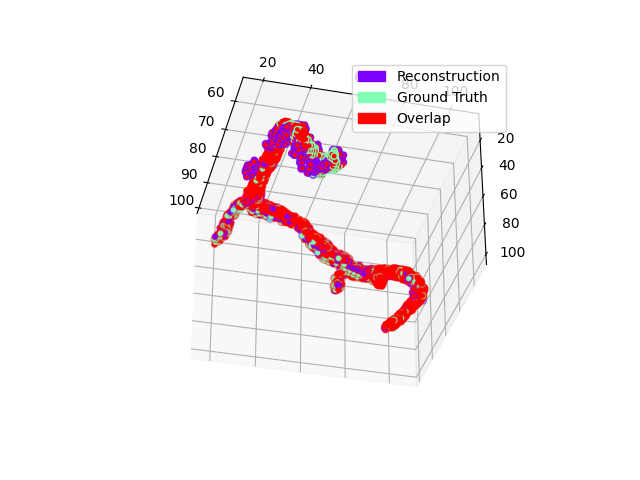}
     \end{subfigure}
     \hfill
	\begin{subfigure}[h]{0.17\textwidth}
         \centering
         \includegraphics[width=\textwidth]{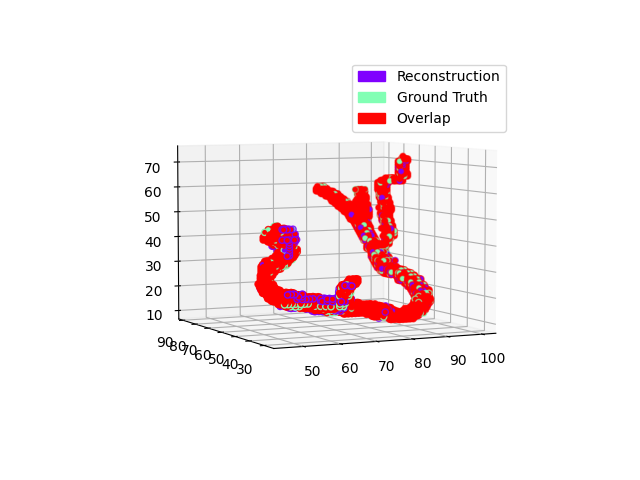}
     \end{subfigure}
     \hfill
	\begin{subfigure}[h]{0.17\textwidth}
         \centering
         \includegraphics[width=\textwidth]{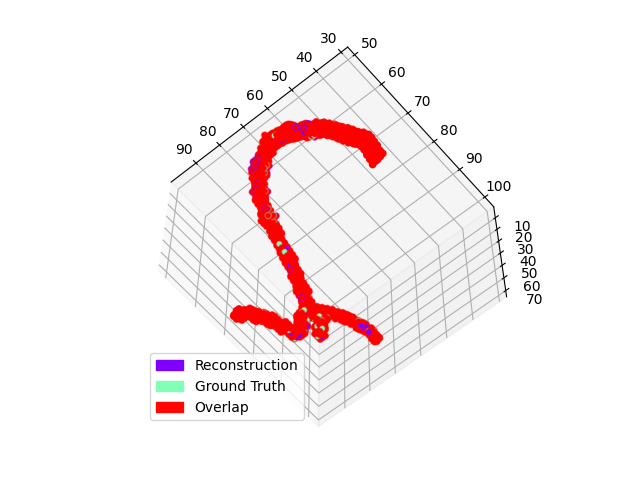}
     \end{subfigure}
     \hfill
	\begin{subfigure}[h]{0.17\textwidth}
         \centering
         \includegraphics[width=\textwidth]{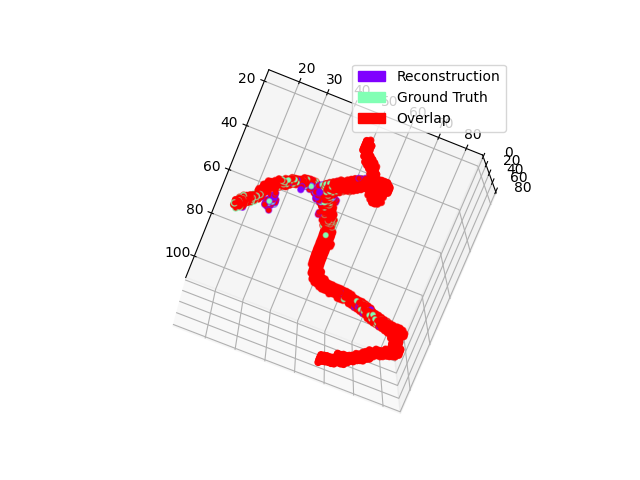}
     \end{subfigure}
     \vfill
	\hspace{0.05cm}3D U-Net\hspace{0.05cm}
	\begin{subfigure}[h]{0.17\textwidth}
         \centering
         \includegraphics[width=\textwidth]{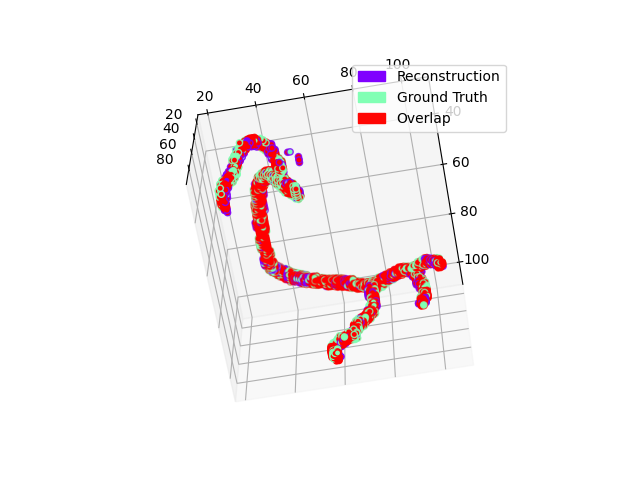}
         \caption*{\hspace{1cm}$R_1$}
     \end{subfigure}
     \hfill
	\begin{subfigure}[h]{0.17\textwidth}
         \centering
         \includegraphics[width=\textwidth]{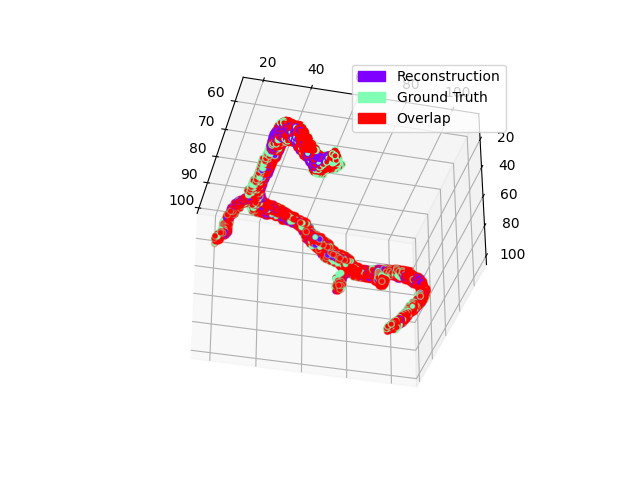}
         \caption*{\hspace{1cm}$R_2$}
     \end{subfigure}
     \hfill
	\begin{subfigure}[h]{0.17\textwidth}
         \centering
         \includegraphics[width=\textwidth]{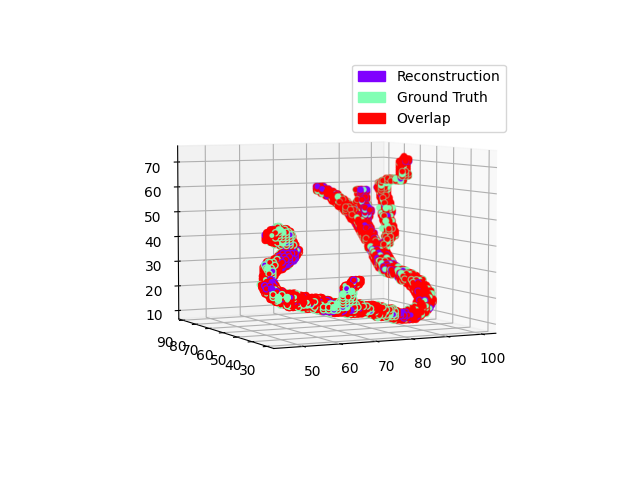}
         \caption*{\hspace{1cm}$R_3$}
     \end{subfigure}
     \hfill
	\begin{subfigure}[h]{0.17\textwidth}
         \centering
         \includegraphics[width=\textwidth]{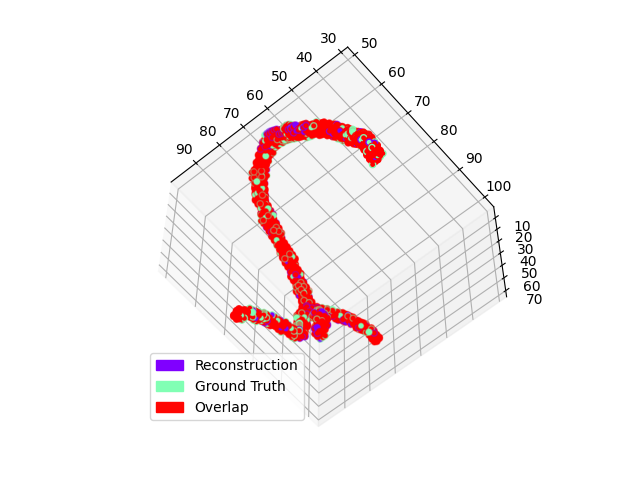}
         \caption*{\hspace{1cm}$R_4$}
     \end{subfigure}
     \hfill
	\begin{subfigure}[h]{0.17\textwidth}
         \centering
         \includegraphics[width=\textwidth]{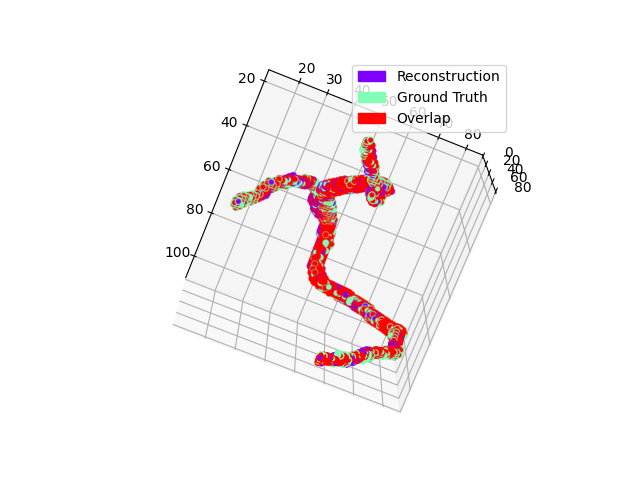}
         \caption*{\hspace{1cm}$R_5$}
     \end{subfigure}
    \caption{Five 3D RCA reconstruction results compared with the corresponding ground truth in the same 3D space. From left to right: five RCA data points $R_{1,2,3,4,5}$. From top to bottom: the comparison results from our NeCA model, NeCA (90$\degree$), and 3D U-Net model. The purple colour represents the reconstruction results; green represents the ground truth; and red shows the overlap between them.}
        \label{visual_combine_rca_1}
\end{figure}

\paragraph{Comparison Between 3D Reconstruction and Ground Truth}
We additionally compare the 3D RCA reconstruction results using the NeCA, NeCA (90$\degree$), and 3D U-Net model with the corresponding ground truth in the same 3D space, as illustrated in Figure~\ref{visual_combine_rca_1}. These figures show that our NeCA model demonstrates better reconstruction overlap than the 3D U-Net model.

\subsubsection{LAD Dataset}
\begin{figure}[!h]
     \hspace{0.3cm}NeCA\hspace{0.3cm}
     \begin{subfigure}[h]{0.17\textwidth}
         \centering
         \includegraphics[width=\textwidth]{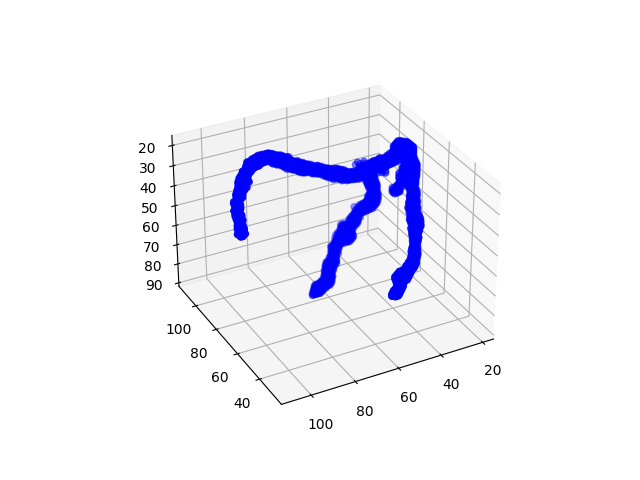}
     \end{subfigure}
     \hfill
	\begin{subfigure}[h]{0.17\textwidth}
         \centering
         \includegraphics[width=\textwidth]{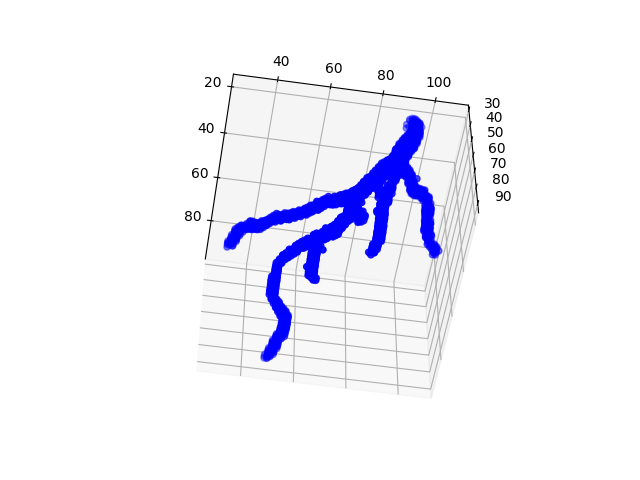}
     \end{subfigure}
     \hfill
	\begin{subfigure}[h]{0.17\textwidth}
         \centering
         \includegraphics[width=\textwidth]{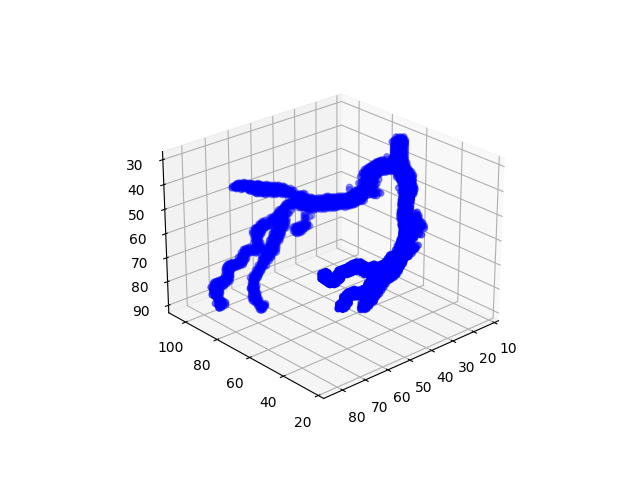}
     \end{subfigure}
     \hfill
	\begin{subfigure}[h]{0.17\textwidth}
         \centering
         \includegraphics[width=\textwidth]{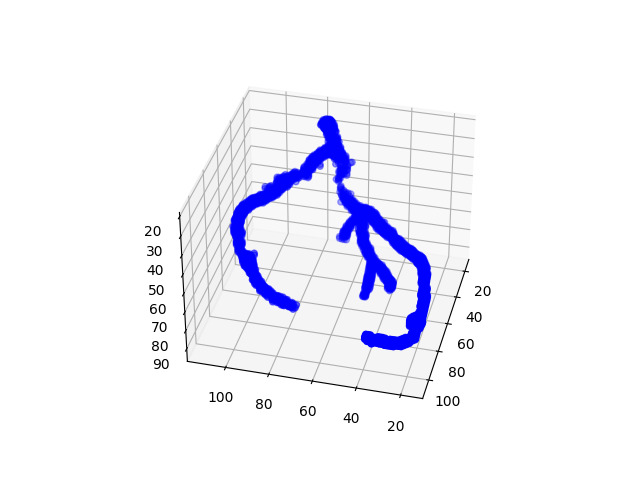}
     \end{subfigure}
     \hfill
	\begin{subfigure}[h]{0.17\textwidth}
         \centering
         \includegraphics[width=\textwidth]{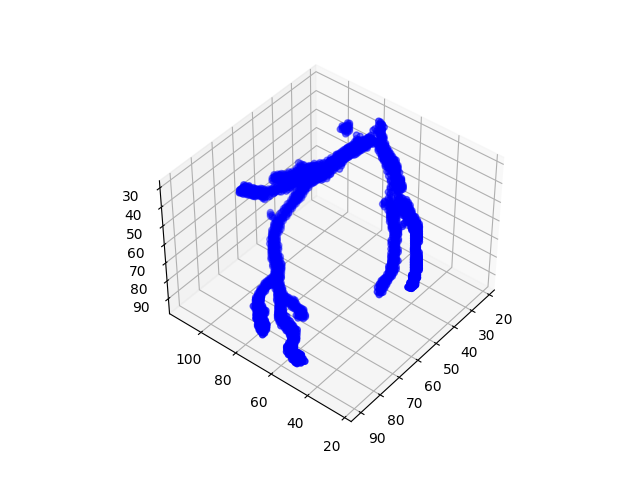}
     \end{subfigure}
     \vfill
	NeCA ($90\degree$)\hspace{-0.11cm}
	\begin{subfigure}[h]{0.17\textwidth}
         \centering
         \includegraphics[width=\textwidth]{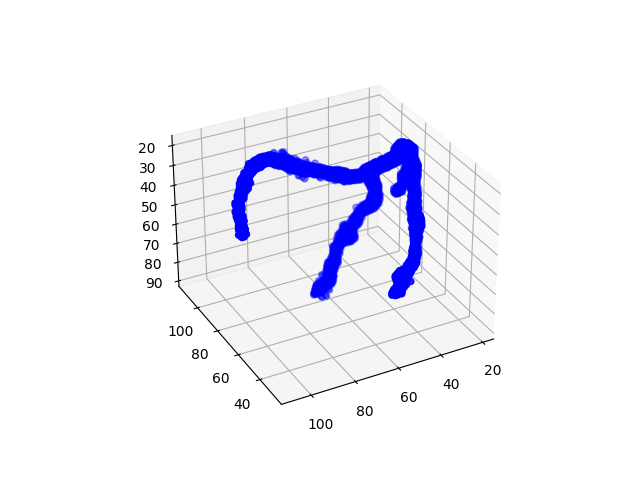}
     \end{subfigure}
     \hfill
	\begin{subfigure}[h]{0.17\textwidth}
         \centering
         \includegraphics[width=\textwidth]{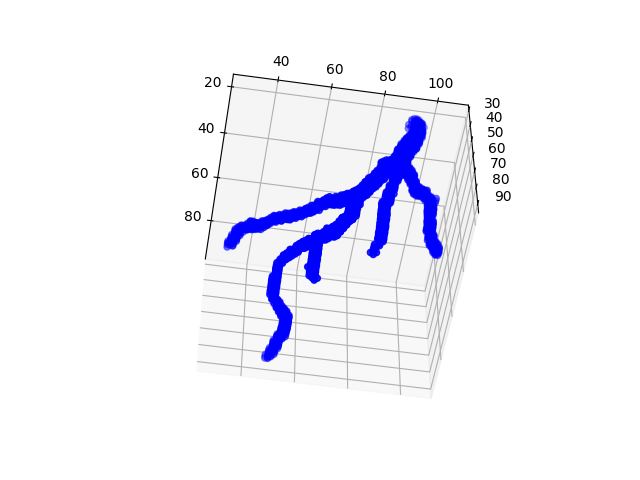}
     \end{subfigure}
     \hfill
	\begin{subfigure}[h]{0.17\textwidth}
         \centering
         \includegraphics[width=\textwidth]{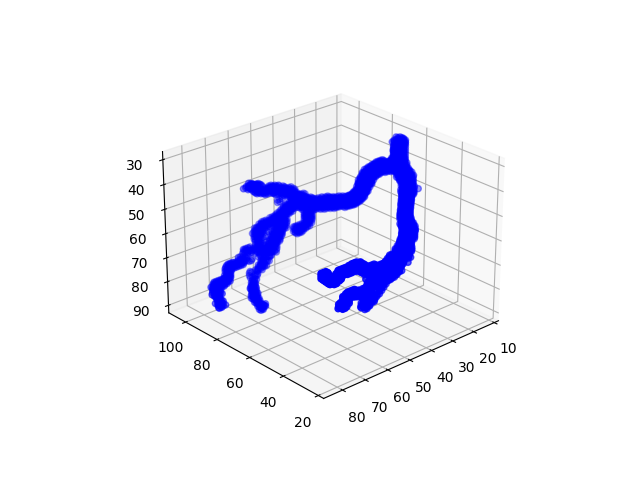}
     \end{subfigure}
     \hfill
	\begin{subfigure}[h]{0.17\textwidth}
         \centering
         \includegraphics[width=\textwidth]{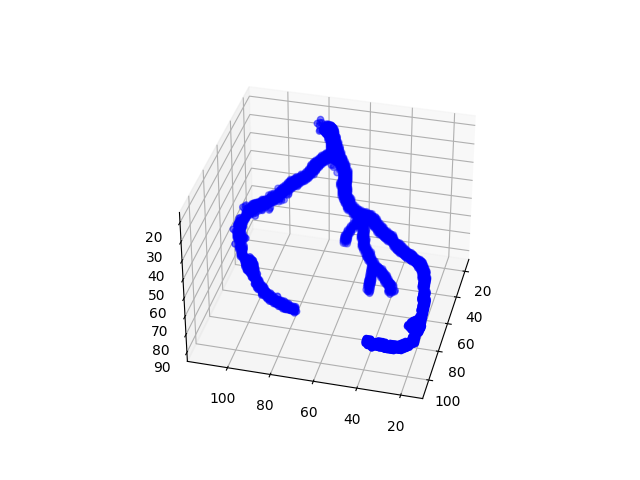}
     \end{subfigure}
     \hfill
	\begin{subfigure}[h]{0.17\textwidth}
         \centering
         \includegraphics[width=\textwidth]{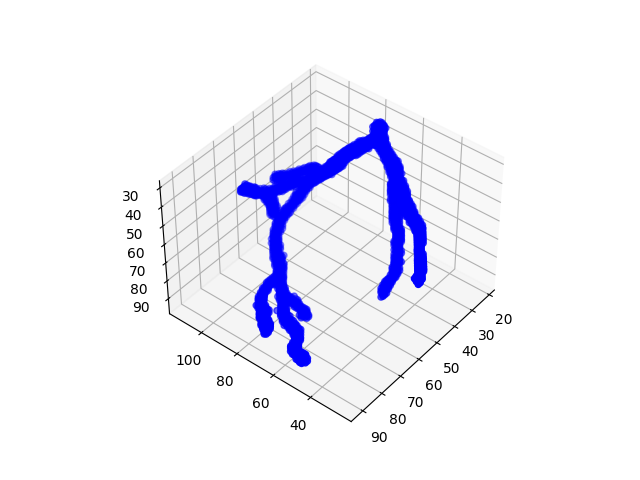}
     \end{subfigure}
     \vfill
     \hspace{0.05cm}3D U-Net\hspace{0.05cm}
     \begin{subfigure}[h]{0.17\textwidth}
         \centering
         \includegraphics[width=\textwidth]{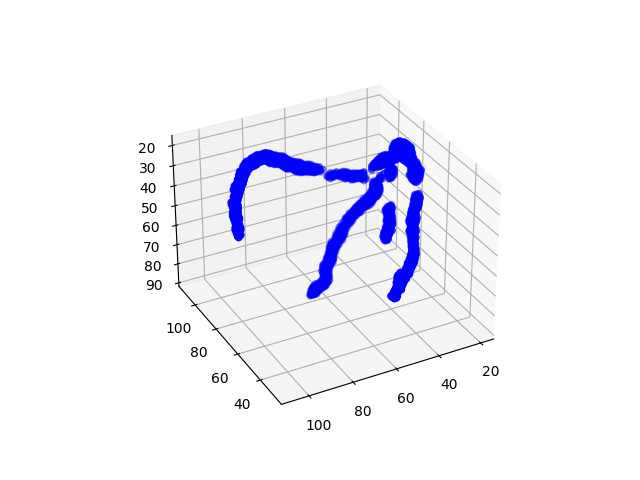}
     \end{subfigure}
     \hfill
	\begin{subfigure}[h]{0.17\textwidth}
         \centering
         \includegraphics[width=\textwidth]{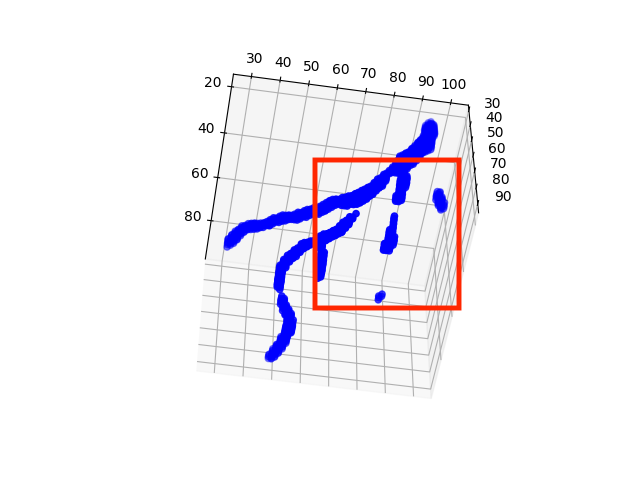}
     \end{subfigure}
     \hfill
	\begin{subfigure}[h]{0.17\textwidth}
         \centering
         \includegraphics[width=\textwidth]{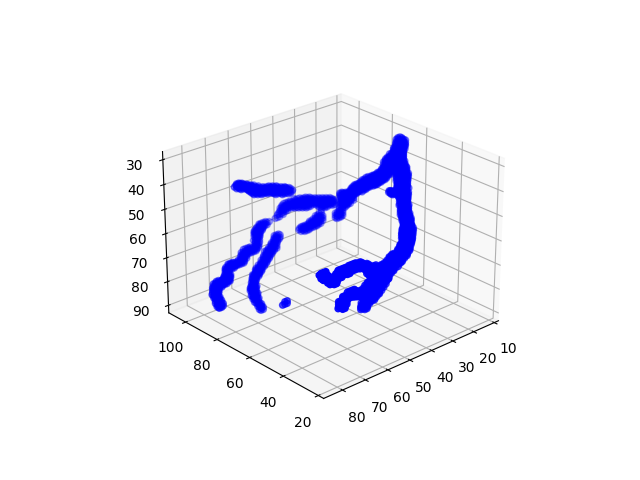}
     \end{subfigure}
     \hfill
	\begin{subfigure}[h]{0.17\textwidth}
         \centering
         \includegraphics[width=\textwidth]{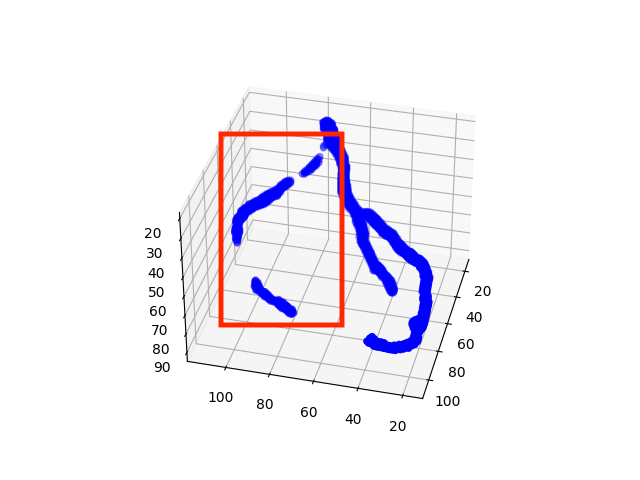}
     \end{subfigure}
     \hfill
	\begin{subfigure}[h]{0.17\textwidth}
         \centering
         \includegraphics[width=\textwidth]{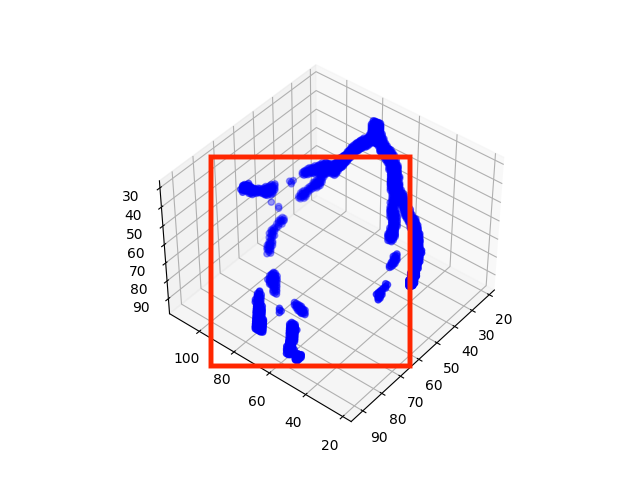}
     \end{subfigure}
     \vfill
     \hspace{0.55cm}GT\hspace{0.55cm}
 	\begin{subfigure}[h]{0.17\textwidth}
         \centering
         \includegraphics[width=\textwidth]{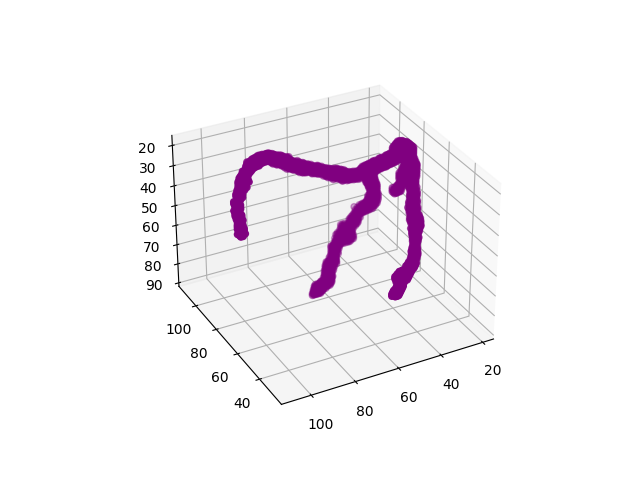}
         \caption*{\hspace{1cm}$L_1$}
     \end{subfigure}
     \hfill
	\begin{subfigure}[h]{0.17\textwidth}
         \centering
         \includegraphics[width=\textwidth]{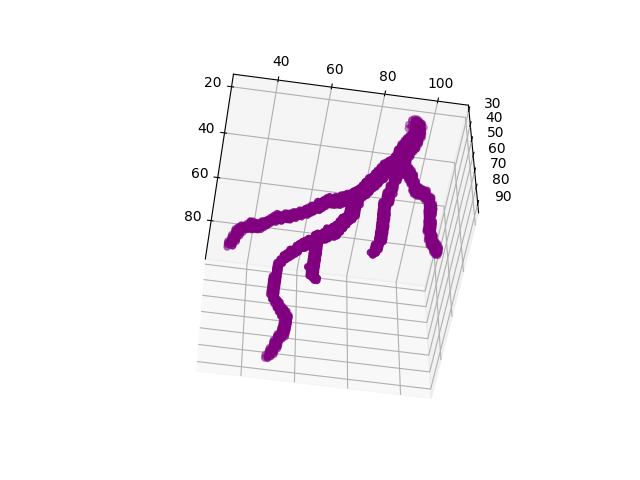}
         \caption*{\hspace{1cm}$L_2$}
     \end{subfigure}
     \hfill
	\begin{subfigure}[h]{0.17\textwidth}
         \centering
         \includegraphics[width=\textwidth]{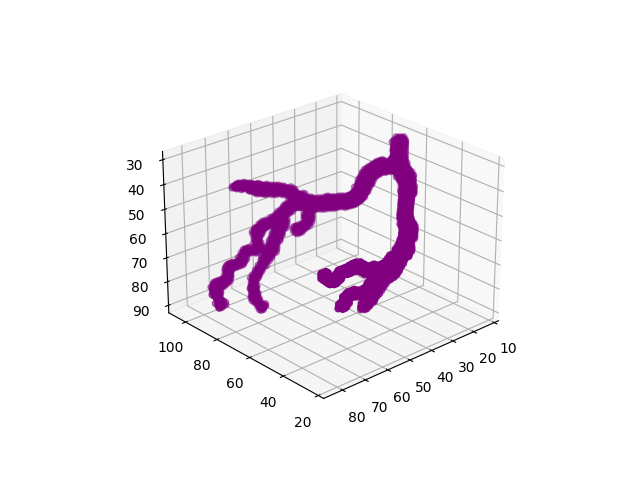}
         \caption*{\hspace{1cm}$L_3$}
     \end{subfigure}
     \hfill
	\begin{subfigure}[h]{0.17\textwidth}
         \centering
         \includegraphics[width=\textwidth]{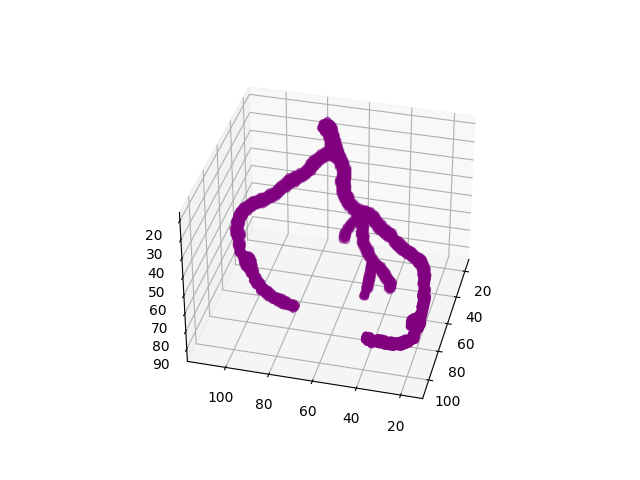}
         \caption*{\hspace{1cm}$L_4$}
     \end{subfigure}
     \hfill
	\begin{subfigure}[h]{0.17\textwidth}
         \centering
         \includegraphics[width=\textwidth]{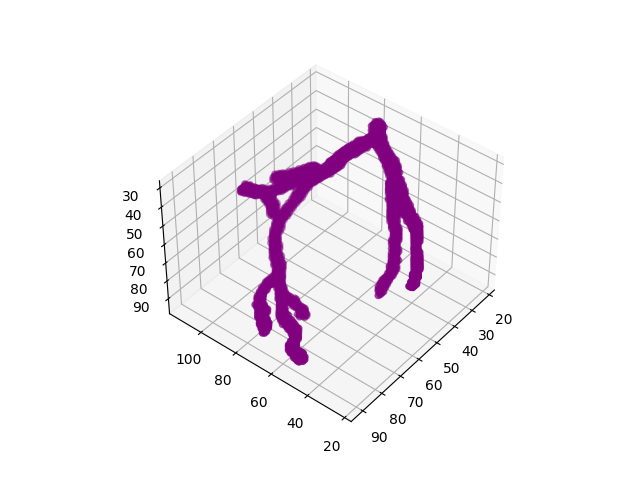}
         \caption*{\hspace{1cm}$L_5$}
     \end{subfigure}
    \caption{Five qualitative 3D LAD reconstruction results. From left to right: five LAD data points $L_{1,2,3,4,5}$. From top to bottom: the reconstruction results using our NeCA model, NeCA (90$\degree$), and 3D U-Net model, along with the corresponding ground truth.}
        \label{visual_recon_lad}
\end{figure}

\paragraph{3D Reconstruction Results}
We show in Figure~\ref{visual_recon_lad} five 3D LAD reconstruction results using our NeCA model, NeCA (90$\degree$), and the 3D U-Net model, with the corresponding ground truth. From the results, we can observe that our NeCA model successfully reconstructs the vasculature of LAD in all five cases. On the other hand, the 3D U-Net model fails to reconstruct some branches in $L_{2,4,5}$ and loses vessel connectivity, as presented in red boxes.

\begin{figure}[!h]
     \hspace{0.3cm}NeCA\hspace{0.3cm}
     \begin{subfigure}[h]{0.17\textwidth}
         \centering
         \includegraphics[width=\textwidth]{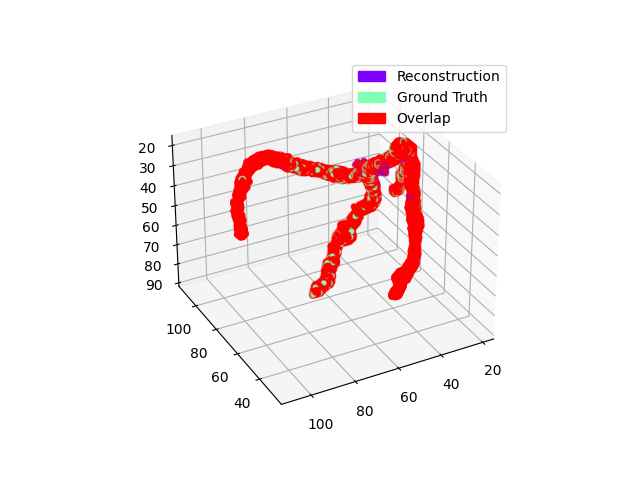}
     \end{subfigure}
     \hfill
	\begin{subfigure}[h]{0.17\textwidth}
         \centering
         \includegraphics[width=\textwidth]{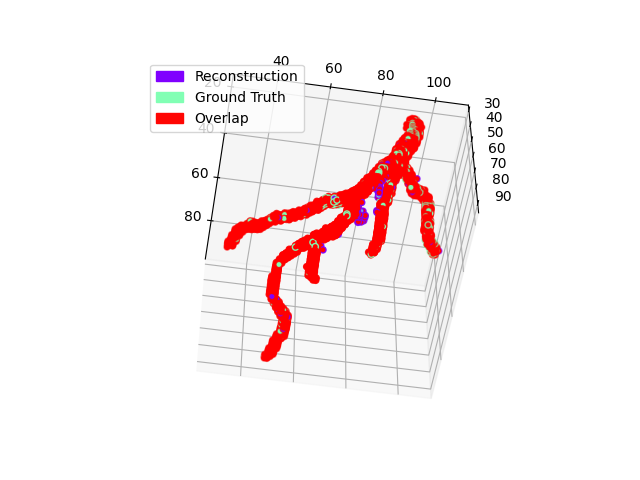}
     \end{subfigure}
     \hfill
	\begin{subfigure}[h]{0.17\textwidth}
         \centering
         \includegraphics[width=\textwidth]{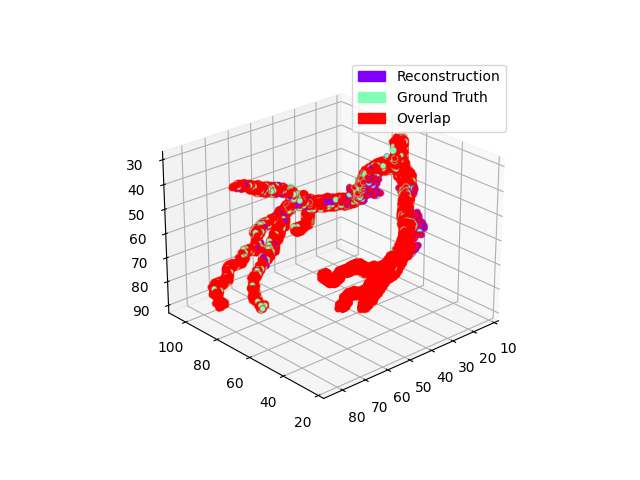}
     \end{subfigure}
     \hfill
	\begin{subfigure}[h]{0.17\textwidth}
         \centering
         \includegraphics[width=\textwidth]{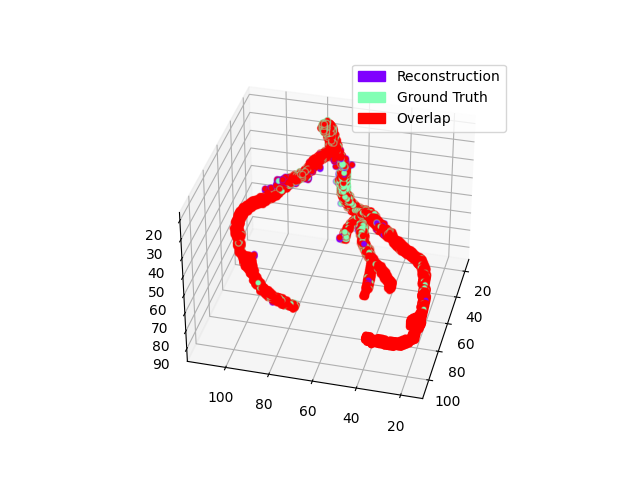}
     \end{subfigure}
     \hfill
	\begin{subfigure}[h]{0.17\textwidth}
         \centering
         \includegraphics[width=\textwidth]{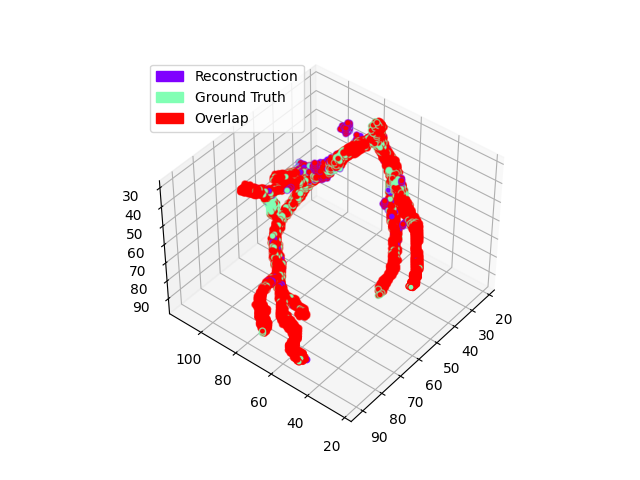}
     \end{subfigure}
     \vfill
     NeCA ($90\degree$)\hspace{-0.11cm}
	\begin{subfigure}[h]{0.17\textwidth}
         \centering
         \includegraphics[width=\textwidth]{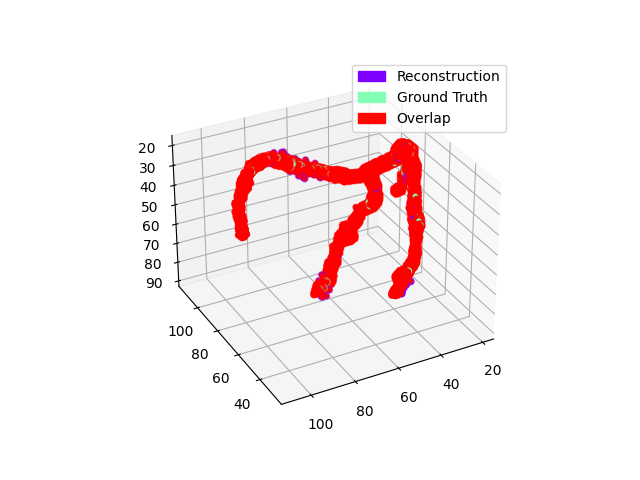}
     \end{subfigure}
     \hfill
	\begin{subfigure}[h]{0.17\textwidth}
         \centering
         \includegraphics[width=\textwidth]{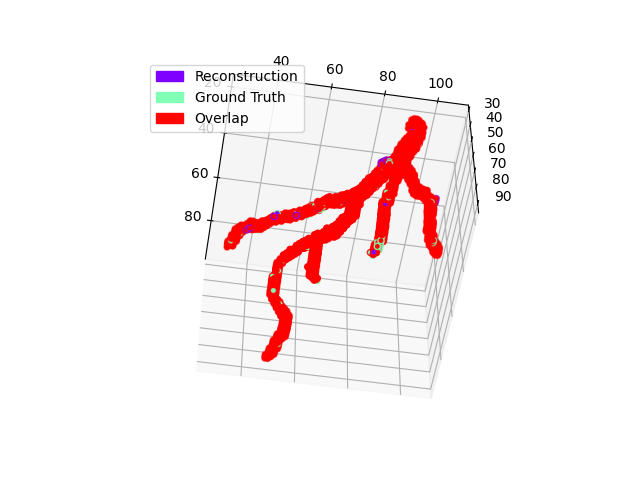}
     \end{subfigure}
     \hfill
	\begin{subfigure}[h]{0.17\textwidth}
         \centering
         \includegraphics[width=\textwidth]{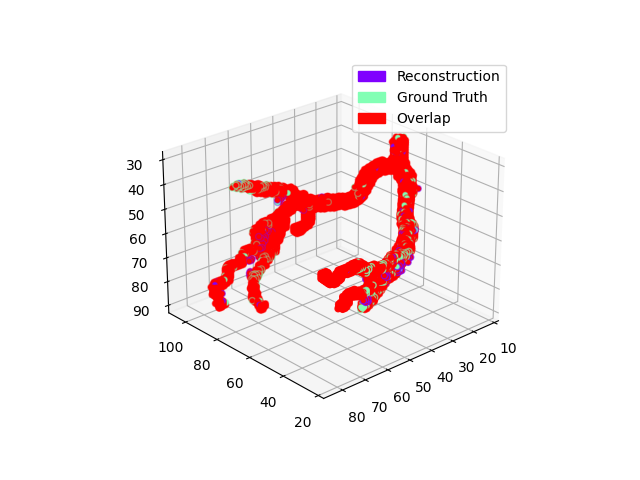}
     \end{subfigure}
     \hfill
	\begin{subfigure}[h]{0.17\textwidth}
         \centering
         \includegraphics[width=\textwidth]{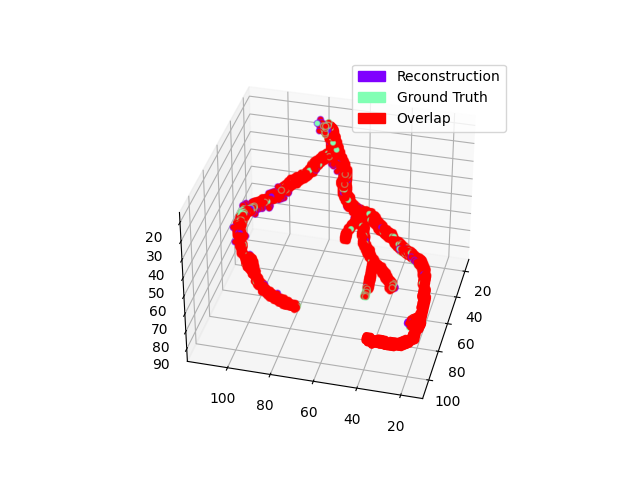}
     \end{subfigure}
     \hfill
	 \begin{subfigure}[h]{0.17\textwidth}
         \centering
         \includegraphics[width=\textwidth]{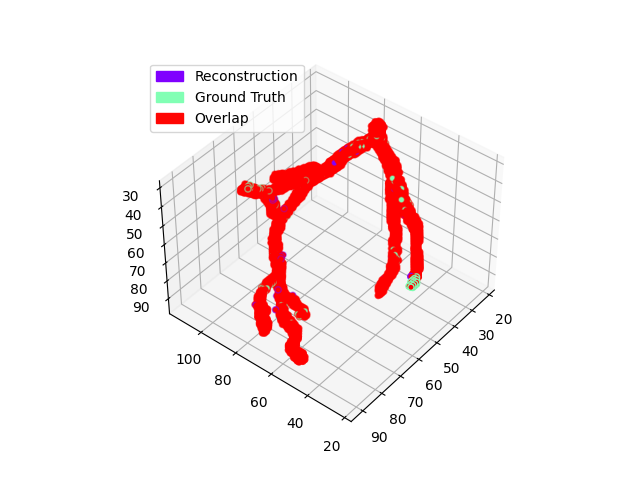}
     \end{subfigure}
     \vfill
     \hspace{0.05cm}3D U-Net\hspace{0.05cm}
	\begin{subfigure}[h]{0.17\textwidth}
         \centering
         \includegraphics[width=\textwidth]{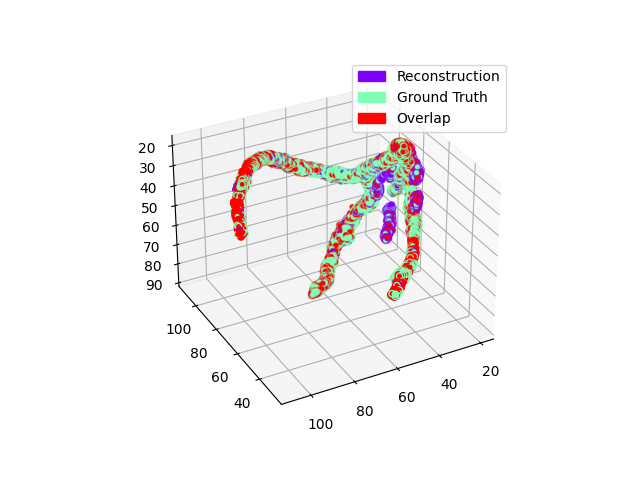}
         \caption*{\hspace{1cm}$L_1$}
     \end{subfigure}
     \hfill
	\begin{subfigure}[h]{0.17\textwidth}
         \centering
         \includegraphics[width=\textwidth]{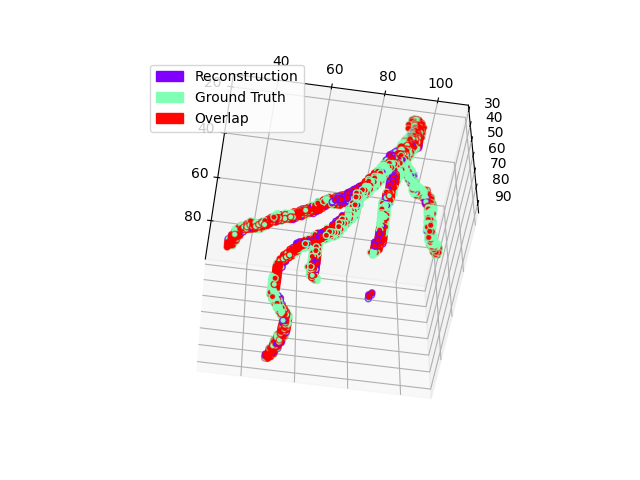}
         \caption*{\hspace{1cm}$L_2$}
     \end{subfigure}
     \hfill
	\begin{subfigure}[h]{0.17\textwidth}
         \centering
         \includegraphics[width=\textwidth]{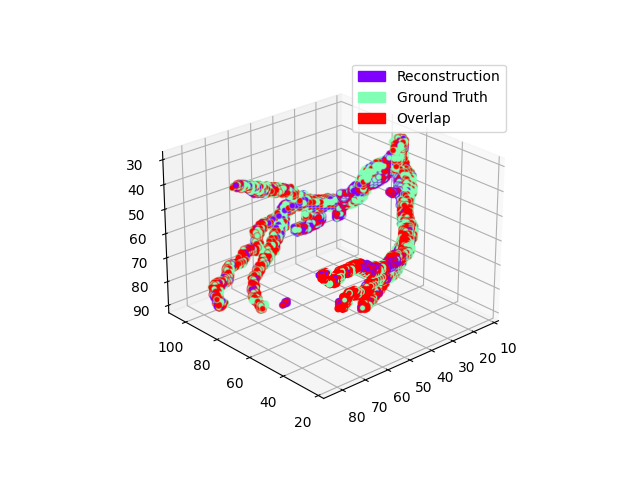}
         \caption*{\hspace{1cm}$L_3$}
     \end{subfigure}
     \hfill
	\begin{subfigure}[h]{0.17\textwidth}
         \centering
         \includegraphics[width=\textwidth]{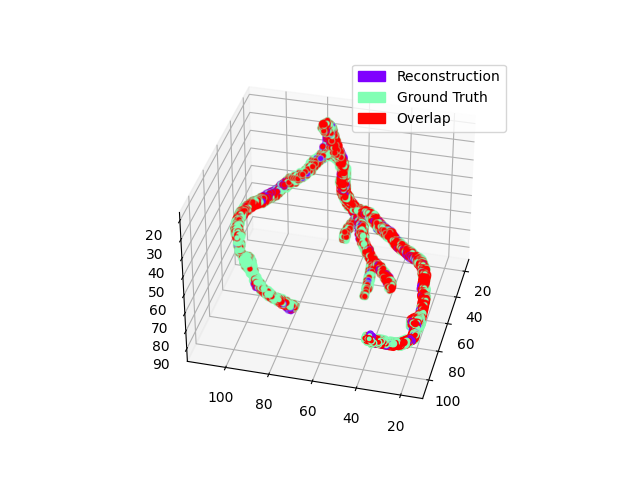}
         \caption*{\hspace{1cm}$L_4$}
     \end{subfigure}
     \hfill
           \begin{subfigure}[h]{0.17\textwidth}
         \centering
         \includegraphics[width=\textwidth]{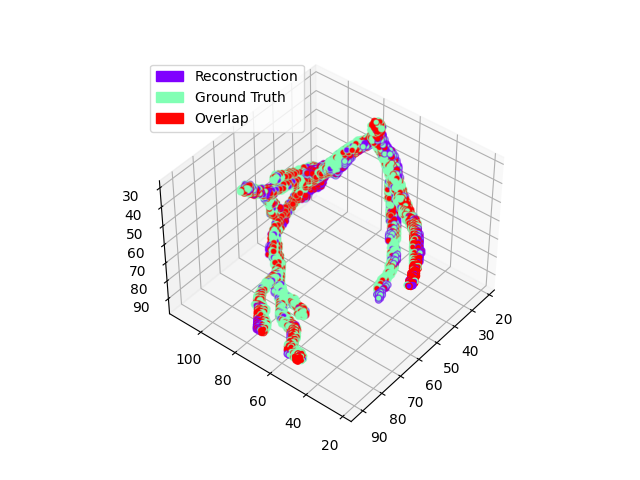}
         \caption*{\hspace{1cm}$L_5$}
     \end{subfigure}
    \caption{Five 3D LAD reconstruction results compared with the corresponding ground truth in the same 3D space. From left to right: five LAD data points $L_{1,2,3,4,5}$. From top to bottom: the comparison results from our NeCA model, NeCA (90$\degree$), and 3D U-Net model. The purple colour represents the reconstruction results; green represents the ground truth; and red shows the overlap between them.}
        \label{visual_combine_lad_1}
\end{figure}

\paragraph{Comparison Between 3D Reconstruction and Ground Truth}
We also compare in Figure~\ref{visual_combine_lad_1} the five 3D LAD reconstruction results using NeCA, NeCA (90$\degree$), and the 3D U-Net models with the corresponding ground truth in the same 3D space. The results show similar performance to the RCA dataset; our NeCA model demonstrates better reconstruction overlap than the 3D U-Net model.

\section{Discussions and Conclusion}
Our evaluation on both the RCA and LAD datasets demonstrates that the NeCA model performs better than the supervised 3D U-Net model in terms of five metrics: \emph{Dice}, \emph{IoU}, \emph{reError}, \emph{CD$_{\ell_2}$}, and \emph{reMSE}. The NeCA model performs statistically significantly better than 3D U-Net model in four metrics for the RCA dataset and five metrics for the LAD dataset out of a total of six metrics. This indicates that our self-supervised learning model, where neither 3D ground truth for supervision nor large training datasets are required, is better than the supervised 3D U-Net model in 3D coronary tree reconstruction from only two projections. It is also demonstrated qualitatively in Section~\ref{qualitative_results} that our NeCA model presents good vasculature reconstruction. In addition, due to the intrinsic properties of our model, we do not need to train two models for RCA and LAD separately, and as a result, it has significant potential to generalise to other tasks.

\par Our model optimised with two orthogonal projections (NeCA (90$\degree$)) shows consistently better performance than our model with two clinical-angle projections (Table~\ref{lad_numerical}), since two orthogonal projections usually contain more feature coverage and less overlapped redundant information (Figure~\ref{projections_example}). However, in real clinics such as cardiac catheterization laboratories, projections are generally not acquired at orthogonal views, thus necessitating this feature of our NeCA model.

\par Our NeCA model contains two trainable components: the hash tables with feature vectors $\mathbf{\Theta}$ from the multiresoultion hash encoder and network parameters $\mathbf{\Phi}$ from the residual MLP. The residual MLP is the backbone of the neural implicit representation, so it cannot be replaced. For the multiresolution hash encoder to encode the coordinates, there are alternative encoders available, such as a frequency encoder, which is not learnable. We have tested the coordinate encoder where we have replaced our multiresolution encoder with a frequency encoder and used the same projection geometry for validation. According to our experiments, the model could not reconstruct any vessels for every case of the RCA and LAD datasets under 5000 iterations.

\par The supervised 3D U-Net model, once trained, can perform real-time 3D coronary tree reconstruction, while our model takes around one hour to optimise the results with a volume size of $128 \times 128 \times 128$ for 5000 iterations. We have also tested our model to optimise a coronary tree of size $64 \times 64 \times 64$, which takes on average of 11 minutes for reconstruction. Therefore, there is a tradeoff between lower reconstruction time and better reconstruction resolution for our NeCA model. The 3D U-Net model applies a pre-trained model during evaluation, so when reconstructing out-of-distribution data, it may fail to generalise, which is a serious threat during clinical applications, whereas our model is optimised for each individual data points and can generalise well. Hence, there is also a tradeoff between real-time reconstruction and stable performance between the 3D U-Net and our NeCA model.

\par The input cone-beam projections to our NeCA model are based on simulation of X-ray intensity attenuation though the object, i.e., the 3D coronary tree. In our experiments using 3D segmented CCTA data, the attenuation coefficients for the coronary tree are assumed to be uniform as a value of $1$. However, in real scenarios, the actual coefficients vary, usually within a certain range due to different vessel conditions. Moreover, blood and contrast injected in the vessel contribute to the X-ray attenuation as well as the other tissues and organs in the background. Though the background removal could be solved with automated coronary vessels segmentation \cite{he2022semistudentteacher,He2023Stacom}, the 3D coronary tree reconstruction based on real X-ray projections with contrast injected and different vessel conditions needs to be explored further.

\par In summary, we have proposed a self-supervised deep learning method, NeCA, using neural implicit representation to achieve 3D coronary artery tree reconstruction from only two projections. Our method neither requires 3D ground truth for supervision nor large training datasets and optimises the reconstruction results in an iterative self-supervised fashion with only the projection data of one patient as input. We leverage the advantages of a learnable multiresolution hash encoder \cite{muller2022instant} to allow for efficient feature encoding, residual MLP neural networks as a continuous function to represent the coronary tree in 3D space, and a differentiable projector layer \cite{jonas_adler_2017_249479} to enable self-supervised learning from 2D input projections. We use a public CCTA dataset \cite{ZENG2023102287} containing both RCA and LAD data to validate our model's feasibility on the task based on six quantitative metrics, and we perform a thorough evaluation. The results demonstrate that our proposed NeCA model achieves promising performance in both vessel topology preservation and branch-connectivity maintenance compared to the supervised 3D U-Net model. Our proposed model also has a high possibility to generalise to other clinical tasks where the ground truth is usually unavailable and hard to acquire.

\bibliographystyle{IEEEtran}
\bibliography{refs}

\end{document}